\documentclass[12pt]{article}

\usepackage[utf8]{inputenc}
\usepackage{multirow}
\usepackage{enumitem}
\usepackage{fullpage}
\usepackage{hyperref}
\usepackage{graphicx}
\usepackage{lipsum}
\usepackage[sort&compress,numbers]{natbib}
\usepackage{hyperref}
\hypersetup{hidelinks}
\usepackage[total={6.5in,9in},top=1in,headsep=0.1in,headheight=1in]{geometry}
\usepackage{amsmath,amssymb}
\usepackage[dvipsnames]{xcolor}
\usepackage{siunitx}
\usepackage[normalem]{ulem}
\usepackage{lineno}
\usepackage{array}
\usepackage{makecell}

\usepackage{aas_macros}

\DeclareRobustCommand{\okina}{%
  \raisebox{\dimexpr\fontcharht\font`A-\height}{%
    \scalebox{0.8}{`}%
  }%
}

\newcommand\snowmass{
\begin{center}
  \rule[-0.2in]{\hsize}{0.01in}\\
  \rule{\hsize}{0.01in}\\
  \vskip 0.1in
  Submitted to the Proceedings of the US Community Study\\ 
  on the Future of Particle Physics (Snowmass 2021)\\
  \rule{\hsize}{0.01in}\\
  \rule[+0.2in]{\hsize}{0.01in}\\[-2em]
\end{center}
}

\usepackage[firstpage=true]{background}
\backgroundsetup{contents={\parbox{6.5in}{\snowmass}}, scale=1,placement=top,opacity=1,color=black,position={3.25in,1.2in}}

\usepackage{fancyhdr}
\fancypagestyle{plain}{%
  \fancyhf{}%
  \fancyhead[C]{}
  \fancyfoot[C]{\thepage}
}

\fancypagestyle{empty}{%
  \fancyhf{}%
  \fancyhead[C]{{\it Dark Matter In Extreme Astrophysical Environments}}
  \fancyfoot[C]{\thepage}
}
\title{
Dark Matter In Extreme Astrophysical Environments}
\date{}

\usepackage{authblk}

\author[1]{Masha Baryakhtar}
\affil[1]{Physics Department, University of Washington, Seattle, WA 98195-1560, USA}

\author[2]{Regina Caputo}
\affil[2]{NASA Goddard Space Flight Center, Greenbelt, MD, USA}

\author[3,4]{Djuna Croon}
\affil[3]{Department of Physics, Durham University, Durham DH1 3LE, UK}
\affil[4]{Institute for Particle Physics Phenomenology, Durham University, Durham DH1 3LE, UK}

\author[5]{Kerstin Perez}
\affil[5]{Massachusetts Institute of Technology, Cambridge, MA 02138, USA }

\author[6]{Emanuele Berti}
\affil[6]{Department of Physics and Astronomy, Johns Hopkins University, 3400 N. Charles Street, Baltimore, Maryland, 21218, USA}

\author[7,8]{Joseph Bramante}
\affil[7]{The Arthur B. McDonald Canadian Astroparticle Physics Research Institute and Department of Physics,
Engineering Physics and Astronomy, Queens University, Kingston, Ontario  K7L 2S8, Canada}
\affil[8]{Perimeter Institute for Theoretical Physics, Waterloo, Ontario N2L 2Y5, Canada}

\author[9]{Malte Buschmann}
\affil[9]{Department of Physics, Princeton University, Princeton, NJ 08544, U.S.A.}

\author[10] {Richard Brito}
\affil[10]{CENTRA, Departamento de F\'{\i}sica, Instituto Superior T\'ecnico -- IST, Universidade de Lisboa -- UL, Avenida Rovisco Pais 1, 1049 Lisboa, Portugal}

\author[11]{Thomas Y. Chen}
\affil[11]{Fu Foundation School of Engineering and Applied Science, Columbia University, New York, NY 10027, USA}

\author[12]{Philippa S. Cole}
\affil[12]{GRAPPA, University of Amsterdam, Science Park 904, 1098XH Amsterdam, The Netherlands }

\author[13,14]{Adam Coogan}
\affil[13]{Département de Physique, Université de Montréal, 1375 Avenue Thérèse-Lavoie-Roux, Montréal, QC H2V 0B3, Canada}
\affil[14]{Mila -- Quebec AI Institute, 6666 St-Urbain, \#200, Montreal, QC, H2S 3H1}

\author[8]{William E.~East}

\author[5]{Joshua W. Foster}

\author[15]{Marios Galanis}
\affil[15]{Stanford Institute for Theoretical Physics, Stanford University, Stanford, CA 94305, USA}

\author[16]{Maurizio Giannotti}
\affil[16]{Department of Chemistry and Physics, Barry University, 11300 NE 2nd Ave., Miami Shores, FL 33161, USA}

\author[17]{Bradley J. Kavanagh}
\affil[17]{Instituto de F\'isica de Cantabria (IFCA, UC-CSIC), Avenida de
Los Castros s/n, 39005 Santander, Spain}

\author[18]{Ranjan Laha}
\affil[18]{Centre for High Energy Physics, Indian Institute of Science, C.\,V.\,Raman Avenue, Bengaluru 560012, India}

\author[19,20]{Rebecca K. Leane}
\affil[19]{SLAC National Accelerator Laboratory, Stanford University, Stanford, CA 94039, USA}
\affil[20]{Kavli Institute for Particle Astrophysics and Cosmology, Stanford University, Stanford, CA 94039, USA}

\author[21,22]{Benjamin~V.~Lehmann}
\affil[21]{Department of Physics, University of California, Santa Cruz, Santa Cruz, CA 95064, USA}
\affil[22]{Santa Cruz Institute for Particle Physics, Santa Cruz, CA 95064, USA}

\author[23]{Gustavo Marques-Tavares}
\affil[23]{Maryland Center for Fundamental Physics, University of Maryland, College Park, MD 20742}

\author[4,24]{Jamie McDonald}
\affil[24]{Centre for Cosmology, Particle Physics and Phenomenology,
Université Catholique de Louvain,
Chemin du cyclotron 2,
Louvain-la-Neuve B-1348, Belgium}

\author[5,25]{Ken~K.~Y.~Ng}
\affil[25]{LIGO Laboratory, Massachusetts Institute of Technology, Cambridge, Massachusetts 02139, USA}

\author[26]{Nirmal Raj}
\affil[26]{TRIUMF, 4004 Wesbrook Mall, Vancouver, BC V6T 2A3, Canada}

\author[27]{Laura Sagunski}
\affil[27]{Institute for Theoretical Physics, Goethe University, 60438 Frankfurt am Main, Germany}

\author[28]{Jeremy Sakstein}
\affil[28]{Department of Physics \& Astronomy, University of Hawai\okina i, Watanabe Hall, 2505 Correa Road, Honolulu, HI, 96822, USA}

\author[29,30,42,43]{B.S. Sathyaprakash}
\affil[29]{Institute for Gravitation and the Cosmos, The Pennsylvania State University, University Park, PA 16802, USA}
\affil[30]{Department of Physics, The Pennsylvania State University, University Park, PA, 16802, USA}

\author[29,30]{Sarah Shandera}

\author[7,8,31]{Nils Siemonsen}
\affil[31]{Department of Physics and Astronomy, University of Waterloo, Waterloo, ON N2L 3G1, Canada}

\author[14]{Olivier Simon}

\author[32]{Kuver Sinha}
\affil[32]{Department of Physics and Astronomy, University of Oklahoma, Norman, OK 73019, USA}

\author[29, 21]{Divya Singh}

\author[33]{Rajeev Singh} 
\affil[33]{Institute of Nuclear Physics Polish Academy of Sciences, PL-31-342 Krak\'ow, Poland}

\author[34]{Chen~Sun}
\affil[34]{School of Physics and Astronomy, Tel-Aviv University, Tel-Aviv 69978, Israel}

\author[35]{Ling Sun}
\affil[35]{OzGrav-ANU, Centre for Gravitational Astrophysics, College of Science,The Australian National University, Australian Capital Territory 2601, Australia}

\author[36]{Volodymyr Takhistov}
\affil[36]{Kavli Institute for the Physics and Mathematics of the Universe (WPI), UTIAS \\The University of Tokyo, Kashiwa, Chiba 277-8583, Japan}

\author[37]{Yu-Dai Tsai}
\affil[37]{Department of Physics and Astronomy, University of California, Irvine, CA 92697-4575, USA}

\author[38]{Edoardo Vitagliano}
\affil[38]{Department of Physics and Astronomy, University of California, Los Angeles, California 90095-1547, USA}

\author[5,24]{Salvatore Vitale}

\author[8,39]{Huan Yang}
\affil[38]{University of Guelph, Guelph, Ontario N2L 3G1, Canada}

\author[40,41]{Jun Zhang}
\affil[40]{Theoretical Physics, Blackett Laboratory, Imperial College London, SW7 2AZ London, UK}
\affil[41]{International Centre for Theoretical Physics Asia-Pacific, Beijing 100190, China}
\affil[42]{Department of Astronomy and Astrophysics, The Pennsylvania State University, University Park, PA, 16802, USA}
\affil[43]{School of Physics and Astronomy, Cardiff University, Cardiff, CF24 3AA, UK}

\begin{document}

\maketitle

\begin{abstract}
Exploring dark matter via observations of extreme astrophysical environments — defined here as heavy compact objects such as white dwarfs, neutron stars, and black holes, as well as supernovae and compact object merger events — 
has been a major field of growth since the last Snowmass process. 
Theoretical work has highlighted the utility of current and near-future observatories to constrain novel dark matter parameter space across the full  mass range. This includes gravitational wave instruments and observatories spanning the electromagnetic spectrum, from radio to gamma-rays. 
While recent searches already provide leading sensitivity to various dark matter models, this work also highlights the need for theoretical astrophysics research to better constrain the properties of these extreme astrophysical systems. The unique potential of these search signatures to probe dark matter adds motivation to proposed next-generation astronomical and gravitational wave instruments.

\end{abstract}

\tableofcontents 

\section{Executive summary}
Astrophysical searches for dark matter (DM) have historically focused on  measuring the cosmic-ray or photon products from the annihilation or decay of a DM particle. 
However, DM interactions could alter the physical processes occurring in the interiors of stars or stellar remnants, the dynamics of black holes, or the mergers of compact objects.
These alterations would imprint characteristic signals in electromagnetic (EM) and gravitational wave (GW) observations.
Exploring DM via observations of these extreme astrophysical environments---defined here as heavy compact objects such as white dwarfs (WDs), neutron stars (NSs), and black holes (BHs), as well as supernovae (SNe) and compact object merger events---has been a major field of growth since the last Snowmass process. 

\begin{figure}[h!]
    \centering
    \includegraphics[width=1\textwidth]{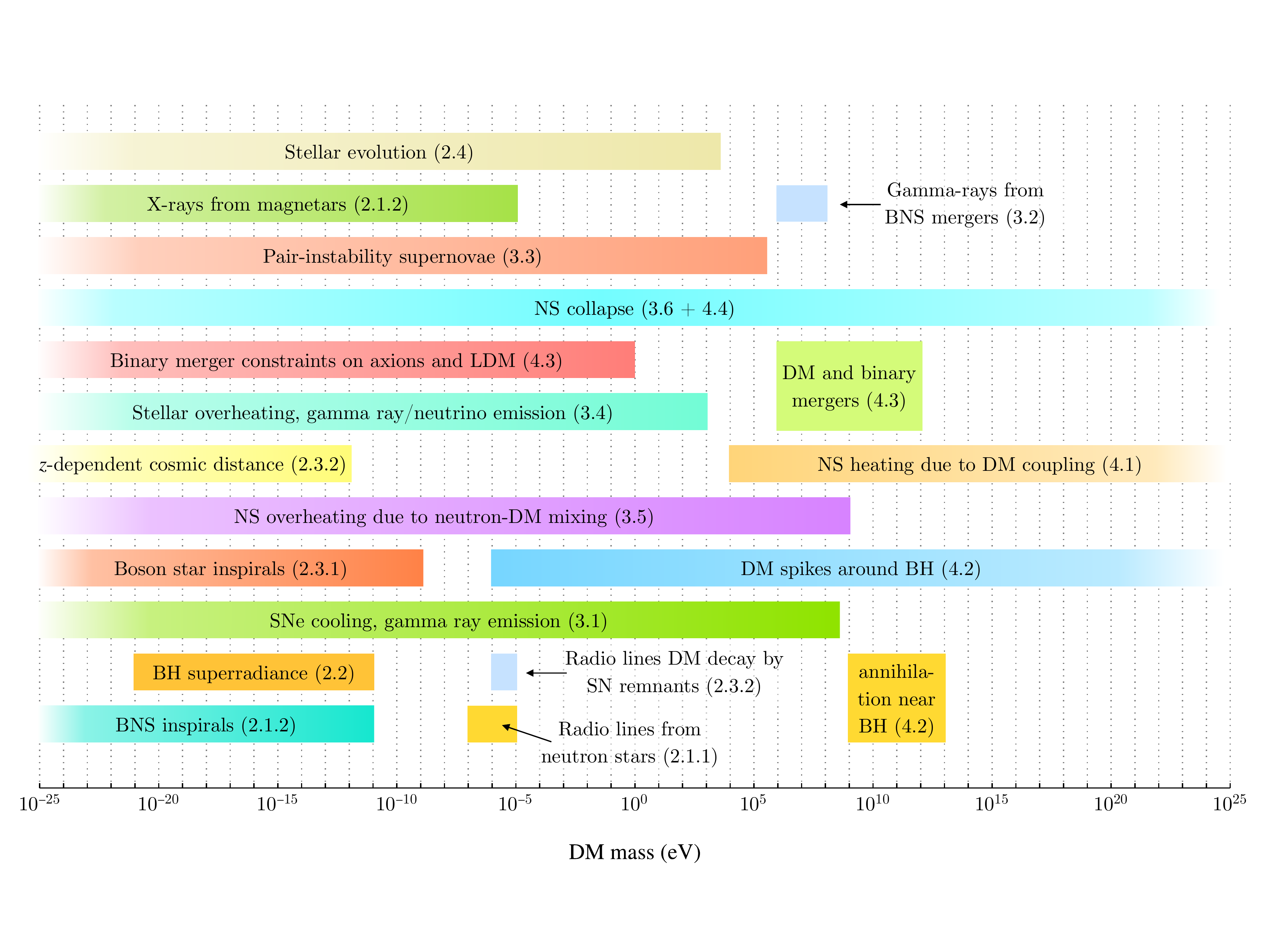}
    \caption{ Summary of the dark matter mass ranges probed by the different methods outlined in this paper. The parenthetical numbers refer to specific sub-sections. 
    }
   \label{fig:key}
\end{figure}

In this white paper, we give an overview of the potential of observations of extreme astrophysical targets to open sensitivity to novel DM parameter space across a broad mass range ($\sim$ 50 orders of magnitude) in the coming decade. Exploiting these opportunities relies on both advances in theoretical work and on current and near-future observatories, including both GW instruments and instruments spanning the full EM spectrum, from radio to gamma-rays. We organize these searches by the DM mass range that is probed: ultralight dark matter (ULDM, $<1$\,keV), light dark matter (LDM, keV--MeV), and heavy dark matter ($\gtrsim$\,GeV). Despite this categorization, we emphasize that many of these probes overlap in mass range, as summarized in Fig.~\ref{fig:key} and Table~\ref{tab:types}.
In addition, we note that in this white paper the parameter space of the DM that is probed does not always saturate the relic abundance; instead, DM is defined as matter which does not interact (appreciably) with Standard Model (SM) matter. 

Extreme astrophysical environments provide unique opportunities to probe ULDM. Ultralight particles can be produced in the hot, dense cores of stars and stellar remnants and affect their evolution.  ULDM---ambient in the environment or produced in the NS itself---can convert in the high magnetic field environment of the NS into radio waves or X-rays that can be searched for in telescopes. In the last decade, new ideas unique to bosonic dark matter have been developed. Specific models of new ULDM can alter the shape of NS waveforms through their coupling to the dense NS matter. Black hole superradiance is a process that can extract energy and angular momentum from rotating astrophysical black holes and place it into bound states of exponentially large numbers of ultralight bosons, as long as the Compton wavelength of the particle is comparable to the BH size. These systems yield signals of coherent gravitational waves as well as BH spin down, which do not depend on particle interactions but only on gravity. Finally, ULDM can form collapsed structures like compact halos and boson stars which can be searched for in gravitational waves or electromagnetic signals.

Opportunities to probe LDM exist from a variety of astrophysical situations: supernovae explosions, the existence of neutron stars, neutron star temperatures, binary neutron star mergers, and black hole population statistics (made possible by gravitational waves from binary inspirals). 
Key observational targets for dark matter in this mass range include observation of gamma rays, neutrinos, and the populations of neutron stars and black holes as observed electromagnetically and via gravitational waves. 
LDM produced in core collapse supernovae can be constrained from limits of their supernova cooling, or lead to visible signals in the x-ray or gamma-ray bands.
LDM produced during a binary neutron star (BNS) merger can lead to a bright transient gamma-ray signal. 
LDM produced in the cores of blue supergiants can affect stellar evolution, ultimately changing black hole population properties including the location of the black hole mass gap. 
LDM scattering and annihilating in exoplantets, brown dwarfs, Population III stars, and stellar remnants can be probed through infrared and optical radiation, and through gamma rays. 
Neutron stars can be heated by LDM via the Auger effect, which is probed by telescopes in the UV, optical and IR ranges of the electromagnetic spectrum. 
Lastly, accumulation of in particular bosonic LDM can lead to the collapse of astrophysical objects.
Most of the signals arise from couplings to  Standard Model photons and fermions.

Compact astrophysical objects such as neutron stars and black holes provide unique test beds for heavy ($>$ GeV) dark matter. 
Dark matter captured by neutron stars and their subsequent heating can be observed by upcoming infrared and radio telescopes. 
High densities of dark matter can collect in spikes around black holes causing enhanced annihilation rates. 
A black hole - compact object binary can form a dark matter spike which can be observed by future space-based gravitational wave observatories. 
Merging compact objects can also give insight into a wide variety of dark sector particles that modify the dynamics of the merger process. 
This includes fifth forces and modifications to gravity. 
Finally, sufficient accumulation of dark matter around a compact object can cause the dark matter particles themselves to collapse into a black hole. 
Upcoming pulsar searches and gravitational wave observatories will be sensitive to this kind of dark matter signature. 
\\

\noindent {\bf Key Opportunities}

In summary, the key opportunities of the coming decade to maximize the sensitivity of these observations to novel dark matter phase space include: 
\begin{itemize}
    \item Collaboration with both observational and theoretical astrophysicists to constrain the standard astrophysical properties of these extreme environments. 
    \item Coordination with the collaborations responsible for the upcoming major observatories: ground- and space-based gravitational wave interferometers, pulsar timing, radio, infrared, X-ray, and neutrino instruments. The goal is two-fold: one, to ensure the performance and capabilities of these future instruments are understood by particle theorists; and two, that observational campaigns and the resulting datasets are optimized as far as possible for the cutting-edge DM search strategies.
    \item Further theoretical development of DM signatures in extreme environments, including theoretical uncertainties and interconnections between observables.
\end{itemize}

\clearpage

\begin{table}[h!]
\footnotesize
\begin{center}
\begin{tabular}{||l | l | l | l | l||} 
 \hline
 Section & Type of DM & signal 
 & mass range & coupling range \\ [0.5ex] 
 \hline
 \ref{sec:NSs}   & \makecell[l]{Product of \\ ALP coupling to \\ nucleons and photons}  &  \makecell[l]{Hard X-rays \\ from  magnetars } & $\lesssim 10^{-5} \mathrm{eV}$  & \makecell[l]{ $G_{aNN}G_{a\gamma\gamma} \, \lesssim \,  10^{-19} \rm GeV^{-2}$  \\ factor $\sim$ 7 improvement \\ with hotter core temperature } \\
 \hline
  \ref{sec:NSs}   & \makecell[l]{Axion DM\\coupling to \\photons}  &  \makecell[l]{Radio lines \\ from  neutron stars } & $\sim 10^{-7} \mathrm{eV}- 10^{-5}\mathrm{eV}$  & \makecell[l]{$G_{a\gamma\gamma} < 10^{-10}-10^{-13} \mathrm{GeV}^{-1}$   \\(improving with \\observing time)} \\
  \hline
 \ref{sec:binary_NSs}   & \makecell[l]{ALP coupling\\ to nucleons}  &  \makecell[l]{GWs from \\ binary neutron \\ star inspirals}  & $\lesssim 10^{-11} \mathrm{eV}$ & \makecell[l]{$1.6 \times 10^{16} <\frac{f_a}{\mathrm{GeV}} < 10^{18}$}  \\
 \hline
  \ref{sec:superradiance}   & Ultralight bosons  &  \makecell[l]{BH superradiance \\ (GWs; BH spin\\ measurements)}  & $ \sim 10^{-21}\,\mathrm{eV} $ - $10^{-11}\,\mathrm{eV}$ & \makecell[l]{gravitational only \\ (potential \\extra signatures \\if other couplings/\\self-interactions \\are present)}  \\
 \hline
   \ref{sec:bosonstars}   & Ultralight bosons  &  \makecell[l]{Boson star inspirals \\ (GWs)}  & $ \sim 10^{-20}\,\mathrm{eV} $ - $10^{-9}\,\mathrm{eV}$ & \makecell[l]{gravitational only \\ (potential \\extra signatures \\if other couplings/\\self-interactions \\are present)}  \\
 \hline
  \ref{sec:axionast}   & \makecell[l]{ALP coupling \\to photons}  &  \makecell[l]{Modification of \\ cosmic distance \\ measurement}  &  $\lesssim 10^{-12}\,\mathrm{eV}$ & \makecell[l]{$G_{a\gamma\gamma} \lesssim  10^{-11}-10^{-12} \mathrm{GeV}^{-1}$}   \\ 
 \hline
 \ref{sec:axionast}   & \makecell[l]{ALP DM coupling \\ to photons}  &  \makecell[l]{Radio from DM\\ decay stimulated\\ by SN remnants}  & $ \sim 10^{-6}\,\mathrm{eV} $ - $10^{-5}\,\mathrm{eV}$ & \makecell[l]{$G_{a\gamma\gamma} \lesssim  10^{-10}-10^{-11} \mathrm{GeV}^{-1}$}  \\ 
 \hline 
  \ref{sec:ULDMstellar}   & \makecell[l]{ALP coupling \\ to photons/electrons}  &  \makecell[l]{Low mass stars \\evolution: Multiple}  & $ \lesssim$ a few keV  & \makecell[l]{$G_{a\gamma\gamma} \sim  10^{-11} \mathrm{GeV}^{-1}$\\ $G_{aee} \sim  10^{-13}$}  \\
 \hline
 \ref{sec:SNe}   & \makecell[l]{DM  coupling\\ to photons or \\SM fermions} & SN: Multiple &  $\lesssim 500 $ MeV & $10^{-12}\, \lesssim G_{a\gamma\gamma} \, \rm GeV \lesssim 5 \times 10^{-5} $  \\ 
  \hline
   \ref{sec:BNS}   & \makecell[l]{DM  coupling to\\ SM charged fermions} & \makecell[l]{Gamma-rays  \\ from BNS merger}  &  $\sim 1$-$100 $  MeV & \makecell[l]{$10^{-12} <\epsilon < 10^{-9} $ \\ (kinetic mixing)}  \\ 
  \hline 
   \ref{sec:BHPS}   & \makecell[l]{LDM coupled\\ to photons/electrons\\ (ALPs, dark photons)} & \makecell[l]{Location of the \\BH mass gap} & $\lesssim 10$ keV / $\lesssim 1$ MeV & \makecell[l]{$\alpha_{26}\gtrsim 1$ (axion-electron)\\
   $g_{10}\gtrsim1$ (axion-photon)\\
   $\varepsilon\gtrsim10^{-7}$ (dark photon) } \\ 
  \hline
  \ref{sec:heatinggamma}   & \makecell[l]{DM with scattering \\ and annihilation \\ processes}  &  \makecell[l]{Stars and planets\\ overheating, or \\ producing gamma \\ rays/neutrinos} & \makecell[l]{$\gtrsim\mathcal{O}$(keV) \\ (depending \\ on object and \\ particle model)} & \makecell[l]{$\sigma_{n\chi} \gtrsim 10^{-47}~{\rm cm}^2$ \\ (depending on object \\ and particle model)} \\  
  \hline
  \ref{subsec:AugerNS}   & \makecell[l]{DM mixing \\with neutrons} & NS overheating & $\lesssim 1.5$~GeV & $ 10^{-17} \leq \epsilon_{nn'}/{\rm eV} \leq 10^{-9}$ \\
  \hline
   \ref{sec:LDMcollapse}   & \makecell[l]{DM coupling \\ to SM fermions }  & NS collapse to BH  & $\sim$ eV-GeV  & $\sigma_{n\chi} \gtrsim 10^{-50}~{\rm cm}^2$ \\
  \hline
\end{tabular}
\end{center}
\end{table}

\clearpage

\begin{table}[]
\footnotesize
\begin{center}
\begin{tabular}{||l | l | l | l | l||} 
 \hline
 Section & Type of DM & signal 
 & mass range & coupling range \\ [0.5ex] 
 \hline
    \ref{subsec:captureNS}   & DM coupling to & NS overheating & 10~keV--10$^{19}$~GeV & $\sigma_{n\chi} \gtrsim 10^{-45}~{\rm cm}^2$ \\
    &  nucleons or leptons & & & \\
      \hline
  \ref{sec:DMspikes}   & \makecell[l]{DM spikes \\ around BH} & GWs  & $10^{-6}\,\mathrm{eV} $ - $10^{62}\,\mathrm{eV}$   & gravitational only  \\
  &  &  & ($\sim 10^{-4}\,M_\odot$) &  \\
  \hline
  \ref{sec:DMspikes}   & DM annihilation   & Gamma rays  & $1\,\mathrm{GeV}$ - $10 \,\mathrm{TeV}$  & thermal relics \\
  & around BH &  &  &  \\
    \hline
\ref{sec:mergers}   & \makecell[l]{Gravitational wave \\signatures of \\ dark sector physics \\ from compact object \\ mergers} & GWs  & \makecell[l]{for axions: \\  $\lesssim$ $\mathcal{O}$(eV).\\ For light dark \\ sector particles: \\  $\lesssim$ 10$^{-10}$ eV. \\ For  DM: \\ $\sim$ MeV -- GeV \\ }  & \makecell[l]{for axions: \\ couplings above \\ reduced Planck mass. \\ For light dark \\ sector particles: \\ $g'$ $\gtrsim$ 10$^{-21}$ \\ For DM:  $\sigma_{n\chi} \gtrsim 10^{-45}~{\rm cm}^2$}. \\
  \hline
  \ref{sec:dmicocollapse}   & \makecell[l]{DM coupling \\ to SM fermions \\ or PBHs }  & \makecell[l]{NS collapse to BH,  \\ (sub)solar-mass BHs, \\ Max pulsar ages, \\  NS/BH mergers, \\ Kilonovae/FRB/GRB, \\Neutrinos from \\ sun/earth, \\ 511 keV signal, \\ r-process \\ nucleosynthesis}  & \makecell[l]{10~keV--10$^{45}$~GeV  \\ PBH: $10^{15}-10^{22}$~g
  }  & \makecell[l]{$\sigma_{n\chi} \gtrsim 10^{-50}~{\rm cm}^2$ \\ gravitational}  \\
 \hline
\end{tabular}
\end{center}
\caption{Summary of the type of dark matter probed, the expected signal, the dark matter mass range, and the coupling range for the methods described in each section.}
\label{tab:types}
\end{table}

\section{Ultralight dark matter (\texorpdfstring{$<$keV}{})}

\subsection{Searches for ultralight bosons with neutron stars}

The extreme astrophysical environments provided by neutron stars provide a powerful means of searching for ultralight bosons, such as axionlike particles (ALPs). Through complementary probes via X-ray, radio, and gravitational wave observation, ALPs can be probed across a wide range of masses with sensitivity comparable to or exceeding laboratory-based searches. In Sec.~\ref{sec:NSs}, we  discuss prospects for detection of $\mu$eV mass axions in radio and sub-neV masses in X-ray. Sub-neV axions may also be detected through their imprints on the gravitational wave signature of neutron star mergers, which we describe in ~\ref{sec:binary_NSs}. Each of these search strategies will gain considerably enhanced sensitivity in the near future through the upcoming Square Kilometer Array in radio, XRISM and ATHENA in X-ray, and LISA, TianQin and Taiji gravitational wave observatories, motivating continued work on theory, observation, and data analysis frontiers to optimize the power of these probes.

\subsubsection{ALP searches with radio emission \& X-rays from neutron stars \label{sec:NSs}}

Radio observations of NSs offer a promising opportunity to detect axion dark matter (DM) through the distinct signature of the conversion of axions to photons in NS magnetospheres. Though the conversion is typically weak, it is exponentially enhanced \cite{Lai:2006af,Battye:2019aco,Foster:2022fxn} by large NS magnetic fields and the degeneracy of the unknown axion mass and the plasma mass of the photon generated by charge density of the NS magnetospheres \cite{Pshirkov:2007st,Hook:2018iia,Huang:2018lxq}. Recent efforts using data from the GBT, Effelsberg, and VLA telescopes have searched for the potentially observable signal, which would appear as a monochromatic radio line at a universal frequency set by the unknown axion mass, yielding leading constraints on axion DM \cite{Foster:2020pgt, Darling:2020plz, Darling:2020uyo, Foster:2022fxn,Battye:2021yue}.

A key advantage to these searches is their sensitivity to axions across a broad range masses of between $10^{-7}$ and $10^{-4}$ eV using existing and upcoming instruments. Moreover, axion searches represent valuable incidental science as any radio observations of locations with large densities in NSs and DM may contain a signal. Finally, these radio probes provide important sensitivity to axions even if the majority of axion DM is gravitationally bound in substructure, which may be expected from generic axion production mechanisms and would reduce laboratory-based sensitivities \cite{Buschmann:2019icd, Vaquero:2018tib, Arvanitaki:2019rax, Eggemeier:2019khm, OHare:2021zrq}. Nonetheless, work remains toward reducing the existing considerable theoretical and observational uncertainties in order to achieve the full power of this search strategy using the upcoming Square Kilometer Array. 
\vspace{1ex}
\begin{enumerate}[topsep=0pt,itemsep=-1ex,partopsep=1ex,parsep=1.5ex]
    \item The precise structure of the NS magnetosphere is an active topic of research in the pulsar community. Current constraints have been derived assuming simple plasma profiles, 
    though more robust predictions and possibly extended sensitivity in axion mass may be realized with state-of-the-art NS models \cite{Philippov:2014mqa, Kalapotharakos:2017bpx}.
    \item The axion conversion signal is jointly determined by the conversion process and the propagation of radio photons through the NS magnetosphere in strong gravitational field of the NS. Accurately calculating this signal and its spectral morphology accounting for all relevant physical effects is a problem of utmost importance and has only begun to be addressed through ray-tracing codes \cite{Leroy:2019ghm, Witte:2021arp, Battye:2021xvt}. Previously unappreciated aspects of axion-photon mixing that may enhance the conversion rate also require careful consideration \cite{Millar:2021gzs}.
    \item The Galactic Center (GC) is dense in NSs and DM so is an optimal environment to detect conversion signals, albeit with large uncertainties, so improved modeling is critical. Other locations of large ambient DM density may also provide new observational targets \cite{Edwards:2019tzf} . 
    \item As dense gravitationally bound substructure may arise in the axion DM scenario, understanding the dynamics of neutron-star-subhalo encounters will provide new prospects of axion detection and require the development of new observational strategies to detect the resulting transient conversion signal \cite{Edwards:2020afl, Buckley:2020fmh, Nurmi:2021xds}. 
\end{enumerate}\vspace{1ex}
Simultaneous with work on these theoretical frontiers, continued observations with GBT, Effelsberg and the VLA alongside new efforts with the Parkes Observatory, the Sardinia Telescope, MeerKAT, the Murchison Widefield Array, and HERA will lead to new constraints on axion dark matter in the immediate future.

 Apart from the radio signals resulting from cold ambient ALPs, another important direction is the study of ALPs produced in the neutron star core subsequently converting to photons in the magnetosphere. ALP production by  nucleon bremsstrahlung processes in the core peaks at energies of a few hundred keV; the ALPs subsequently  escape and  convert to photons, giving rise to emission in the hard X-ray and soft gamma-ray bands. Furthermore, ALPs escaping the core are effectively cooling the neutron star, an effect that can be potentially observed in the luminosity data of old stars (see e.g.~\cite{Beznogov:2018fda,Buschmann:2021juv}).  

The parametric dependence of the ALP emissivity and conversion rate 
indicate that young  magnetars, with high core temperatures and strong magnetic fields, are ideal targets for probing ALPs \cite{Fortin:2018ehg, Fortin:2018aom,  Fortin:2021cog}. Upper limits on the product of the ALP-nucleon and ALP-photon coupling can be derived by minimally demanding that the ALP-induced emission does not exceed the actual experimentally observed emission from a target, regardless of the astrophysical background. Data from NuSTAR, INTEGRAL, and XMM-Newton has been utilized to place constraints on ALPs from  a set of eight magnetars for which hard X-ray data exists \cite{Fortin:2021sst}; published quiescent soft-gamma-ray flux upper limits  obtained with CGRO, COMPTEL and INTEGRAL SPI/IBIS/ISGRI have been used to obtain constraints on ALPs from five magnetars \cite{Lloyd:2020vzs}. Future experiments like AMEGO covering the ``MeV gap" would place better  constraints.  Other systems -- such as magnetic white dwarfs \cite{Dessert:2019sgw} and  the Magnificent Seven \cite{Buschmann:2019pfp, Dessert:2019dos} -- have also been studied recently  in the same framework. Tantalisingly,~\cite{Dessert:2019dos} observed an excess of hard X-Ray emission in several of the Magnificent Seven neutron stars which could be explained by axions~\cite{Buschmann:2019pfp}. There is no known astrophysical explanation for this excess but its origin can be tested by future NuSTAR observations.

Challenges in modeling  the magnetar heating mechanism makes it difficult to precisely know the temperature of the magnetar core, which is the quantity that most influences the emissivity. 
Varying benchmark choices while keeping the core temperature fixed  typically results in less than a factor of three uncertainty in the ALP spectrum and subsequently the upper limits on the product of ALP couplings \cite{Fortin:2021sst}. 
  
 On  the observational side, improved limits can be obtained by incorporating  the astrophysical background coming from resonant Compton scattering and  discriminating the ALP-induced emission. The morphology of the resonant Compton scattering background depends on several factors, such as the electron Lorentz factor, surface temperature, and magnetar viewing angle. While the magnetosphere is opaque to hard X-ray and gamma-ray emission due to magnetic photon splitting and pair production, ALP-induced emission occurs  near the radius of conversion which is typically several thousand kilometers from the surface, where such effects are expected to be reduced. Moreover, a  comparison of phase-resolved spectra is expected to be particularly discriminating, as is a comparison of polarization patterns. Experiments like X-Calibur and IXPE targeting hard X-ray polarimetry of neutron stars are important in this regard \cite{Krawczynski:2019ofl}.

\subsubsection{Binary neutron star searches for axionlike particles \label{sec:binary_NSs}}

In the dense environment of a neutron star, light axions could receive finite density corrections to their potential if they couple to nuclear matter in a similar way as the QCD axion. For axions with a decay constant below $10^{18} {\rm GeV}$, and lighter than the QCD axion, the finite density corrections can cause a phase transition of the axion field, endowing the neutron star with an axion charge \cite{Hook:2017psm}. In this case, the axion field mediates an additional force between neutron stars, with a strength that could be as strong as gravity. The axion force can be either attractive or repulsive, depending on whether the axion field values are of the same or opposite sign on the surfaces of the two neutron stars; in addition the axion field may radiate axion waves during binary neutron star coalescence.

Changes to NS inspirals in the presence of the axion field have been investigated in Refs.~\cite{Sagunski:2017nzb,Huang:2018pbu};  effects on the inspiral waveform from the axion field were calculated to the first post-Newtonian order. Analysis of gravitational waves from the binary neutron star inspiral GW170817  excludes axions with masses below $10^{-11} {\rm eV}$ and decay constants ranging from $1.6 \times 10^{16} {\rm GeV}$ to $10^{18} {\rm GeV}$ at the 3$\sigma$ confidence level \cite{Zhang:2021mks}.

The parameter space probed by binary neutron star inspirals can be improved with future theoretical work and more observations. The analysis in Ref.~\cite{Zhang:2021mks} does not consider the induced charge effect~\cite{Huang:2018pbu}, which could become important at the late inspirals for axions with small masses. Taking into account the induced charge effect could potentially extend the detectable region to $10^{16} {\rm GeV} < f_a < 10^{18} {\rm GeV}$ for $m_a \le 10^{-14} {\rm eV}$~\cite{Zhang:2021mks}. Constraints from binary NS inspirals can also be improved if the SNR of the merger event is enhanced, for example by stacking multiple binary NS merger events or with the next generation of GW detectors, such as Einstein Telescope and Cosmic Explorer.

\subsection{Black hole superradiance searches for ultralight bosons}
\label{sec:superradiance}

Black holes can serve as discovery engines for ultralight bosons~\cite{Arvanitaki:2009fg,Arvanitaki:2010sy}.
Through physics akin to the classical Penrose process, spinning black holes may superradiantly amplify excitations in a surrounding field, shedding some of their energy and angular momentum in the process~\cite{Brito:2015oca}.
The nonzero mass of the boson allows the excitations (particles) to remain confined close to the hole, exponentially driving the field amplification until the black hole has lost enough angular momentum so that the process shuts down.
In this way, rapidly spinning black holes can spontaneously trigger the formation of a macroscopic bound state of the boson field---a boson ``cloud''---which could persist for a long time and lead to rich phenomenology.
Bosons in the cloud arrange themselves in discrete energy levels analogous to those of the hydrogen atom.
The cloud can have a mass up to ${\sim}10\%$~\cite{Herdeiro:2021znw} of the original black hole, resulting in the emission of continuous, quasimonochromatic gravitational waves that eventually drain the cloud and make it disappear.
Before that happens, particle transitions between the energy levels of this ``gravitational atom'' can also produce gravitational waves~\cite{Arvanitaki:2009fg,Arvanitaki:2010sy,Arvanitaki:2014wva,Siemonsen:2019ebd}. 
On the other hand, if the bosons have additional interactions, new processes can affect the dynamics and evolution of the cloud~\cite{Arvanitaki:2010sy,Yoshino:2012kn,Arvanitaki:2014wva,Ikeda:2018nhb,Boskovic:2018lkj,Fukuda:2019ewf,Baryakhtar:2020gao,Omiya:2020vji,Omiya:2022mwv}. 
All of these phenomena lead to observational signatures that could allow us to detect the existence of ultralight bosons, and establish (or constrain) their role as a constituent of dark matter.

The superradiant instability is only triggered when black holes spin fast enough, and is most efficient when the boson's Compton wavelength is comparable to the black hole's horizon radius. Therefore the boson parameter space that can be explored by these means is set by the astrophysical distribution of black-hole masses and spins.
The requirement that the boson wavelength match the horizon radius implies that stellar mass black holes in the $[1, 10^3]\, M_\odot$ range could allow us to probe bosons with masses within $[10^{-11}, 10^{-14}]\, \mathrm{eV}$; on the other hand, supermassive black holes with mass $[10^3, 10^8]\, M_\odot$ could allow us to probe bosons with $[10^{-19}, 10^{-14}]\, \mathrm{eV}$~\cite{Arvanitaki:2009fg,Arvanitaki:2010sy,Arvanitaki:2014wva,Baryakhtar:2017ngi,Brito:2017zvb}.
The former (latter) can be explored by present and future ground-based (space-based) gravitational-wave detectors, giving us access to a part of the ultralight boson parameter space that 
is difficult to probe in other ways.
Besides gravitational-wave observations, mass and spin measurements of black holes inferred through electromagnetic observations, such as the ones coming from stellar mass black holes in X-ray binaries or supermassive black holes as measured through continuum fitting and the K$\alpha$ iron line~\cite{Reynolds:2019uxi}, can also  be used to probe the existence of bosons in similar mass ranges~\cite{Arvanitaki:2010sy,Pani:2012vp,Arvanitaki:2014wva,Baryakhtar:2017ngi,Cardoso:2018tly,Stott:2018opm}.

In the subsections below, we elaborate on the exciting prospect of detecting ultralight bosons through their gravitational-wave signals, or through their imprint on the distribution of observed black-hole masses and spins.
We also discuss open questions in our understanding of superradiant boson clouds around black holes, and the numerical or theoretical avenues to answer them.

\subsubsection{Gravitational wave signals from ultralight boson clouds}

As discussed above, the formation of macroscopic boson clouds around astrophysical black holes would result in the emission of continuous, quasimonochromatic gravitational waves, that slowly dissipate the energy stored in the cloud. In particle physics terms, those gravitational waves arise due to the annihilation of the bosons in the cloud, and provide one of the most exciting signatures from ultralight bosons that can be searched for with current and future gravitational wave detectors~\cite{Arvanitaki:2009fg,Arvanitaki:2010sy,Yoshino:2013ofa,Arvanitaki:2014wva,Baryakhtar:2017ngi,Brito:2017zvb,Brito:2017wnc,Zhu:2020tht,Brito:2020lup}. The typical frequency of the gravitational waves emitted by these clouds falls in the most sensitive bucket of current~\cite{KAGRA:2013rdx} and future Earth-based detectors~\cite{Kalogera:2021bya} for bosonic fields in the range $\sim 10^{-14}$~--~$10^{-11}$~eV, while planned space-borne gravitational wave detectors such as LISA~\cite{LISA:2017pwj} or TianQin~\cite{TianQin:2020hid}, will be sensitive to a complementary mass range $\sim 10^{-19}$~--~$10^{-14}$~eV. The gravitational wave frequency is only very weakly dependent on the spin of the boson field, i.e. whether it is a scalar, vector or even tensor field, however the radiated power, which affects the signal lifetime and the rate at which the frequency changes, can be significantly stronger for vector and tensor fields compared to scalar fields~\cite{Baryakhtar:2017ngi,Siemonsen:2019ebd,Brito:2020lup}. In addition to the boson spin, if sufficiently strong, self-interactions or couplings to other matter fields can also significantly affect the development of the superradiant instability and subsequent gravitational wave emission~\cite{Yoshino:2012kn,Fukuda:2019ewf,Baryakhtar:2020gao,Omiya:2020vji,Rosa:2017ury,Sen:2018cjt,Ikeda:2018nhb,Boskovic:2018lkj}.

Gravitational waves from boson clouds can be searched for by targeting known black holes, such as black hole remnants formed from the merger of compact binary coalescences or known X-ray binaries~\cite{Arvanitaki:2016qwi,Baryakhtar:2017ngi,Isi:2018pzk,Ghosh:2018gaw,Sun:2019mqb,Ng:2020jqd}, or performing blind all-sky searches for signals coming from a population of unknown isolated black holes~\cite{Arvanitaki:2014wva,Baryakhtar:2017ngi,Brito:2017zvb,Brito:2017wnc,DAntonio:2018sff,Palomba:2019vxe,Zhu:2020tht,LIGOScientific:2021jlr}, which is especially promising given that just in the Milky Way we expect up to $\sim 10^8$ isolated black holes~\cite{Zhu:2020tht}. In addition to searching for individual sources, the expectation that a large number of sources too faint to be extracted from noise should exist can also lead to a stochastic gravitational wave background, which has been computed in Refs.~\cite{Brito:2017zvb,Brito:2017wnc,Tsukada:2018mbp,Zhu:2020tht,Tsukada:2020lgt,Yuan:2021ebu}.

Different types of searches for these sources have already started to be carried out with current gravitational wave detectors~\cite{Tsukada:2018mbp,Sun:2019mqb,Palomba:2019vxe,Zhu:2020tht,Tsukada:2020lgt,LIGOScientific:2021jlr}. The lack of a detection so far is already starting to disfavor parts of the parameter space for bosons with masses in the range $\sim 10^{-13}$--$10^{-12}$~eV~\cite{Tsukada:2018mbp,Sun:2019mqb,Palomba:2019vxe,Zhu:2020tht,Tsukada:2020lgt,LIGOScientific:2021jlr}. With the increasing sensitivity of the next-generation of ground-based detectors these constraints are expected to be greatly improved, or even possibly lead to a detection~\cite{Arvanitaki:2016qwi,Baryakhtar:2017ngi,Isi:2018pzk,Ghosh:2018gaw,Yuan:2021ebu}. Jointly with planned space-based detectors, which will be sensitive to gravitational waves at lower frequencies~\cite{Arvanitaki:2014wva,Brito:2017zvb,Brito:2017wnc}, in the next decades gravitational wave observations will potentially constrain or detect ultralight bosons in the whole mass range $\sim 10^{-19}$~--~$10^{-11}$~eV; for further details, see \cite{Brito:2022lmd}.

Besides gravitational waves emitted directly by the cloud, if the black hole-boson cloud system is part of a binary, a wealth of other effects can occur that lead to very distinct and potentially detectable signatures in the gravitational waves emitted by the coalescing binary~\cite{Baumann:2018vus}. Those include signatures induced by the tidal field of the companion object, such as level mixing, ionization, tidal resonances and tidal disruption ~\cite{Baumann:2018vus,Zhang:2018kib,Baumann:2019ztm,Berti:2019wnn,Cardoso:2020hca,Takahashi:2021yhy,Baumann:2021fkf}, non-vanishing tidal Love numbers~\cite{Baumann:2018vus,DeLuca:2021ite}, signatures induced by the multipolar structure of the boson cloud~\cite{Baumann:2018vus,Su:2021dwz} and effects such as the accretion of the cloud by the companion object, dynamical friction, and the impact caused by the self-gravity of the cloud itself~\cite{Barausse:2014tra,Hannuksela:2018izj,Zhang:2019eid,Baumann:2021fkf,Tahura:2022ffs,Traykova:2021dua,Vicente:2022ivh}. All those effects taken together can contribute to a potentially observable change in the gravitational signal emitted by a binary black hole, if one or both black holes in the binary are surrounded by a boson cloud.

\subsubsection{Black hole spindown signals of ultralight bosons}

As discussed above, clouds of ultralight bosons can form around a black hole of appropriate mass if the black hole's spin is large enough. The cloud will gradually reduce the spin of the host black hole, with a timescale that depends on the strength of the bosons self-interaction. If self-interaction is entirely ignored, the fastest-growing cloud can develop in a matter of minutes for stellar-mass black holes.
Ultralight bosons are thus expected to leave a measurable imprint on the spin distribution of black holes old enough that superradiance has had the time to act. Black holes in binaries, such as those with stellar masses detected by ground-based gravitational wave detectors (LIGO, Virgo, KAGRA) are ideal sources to search for ultralight bosons, as they generally inspiral for millions of years before merging in the band of the detectors. That allows for the formation of the first few clouds to reduce the spin (and the mass, though that is a smaller effect) of the host black hole~\cite{Arvanitaki:2014wva,Brito:2017zvb}. 
In general, black holes are not expected to be efficiently spun-up again through accretion of the surrounding material, since the spin-up time scale can be orders of magnitude longer than the time scale of the superradiant instability, even at the Eddington accretion rate~\cite{Gou2011}. As a result, a large number of low-spin black holes in a particular mass range might hint the existence of the boson clouds.

As ground-based detectors will measure hundreds of stellar-mass binary black holes per year, one can thus verify if this population has unusually small spins, which might indicate that they have been spun-down by superradiance \cite{Arvanitaki:2016qwi}. Unfortunately, such measurement is made challenging by the fact that it is partially degenerate with the underlying (and currently unknown) distribution of stellar mass black hole spins at formation originated from astrophysical processes. On the other hand, it is much easier to \emph{exclude} the existence of ultralight bosons in some mass range, if one finds a few fast spinning black holes inconsistent with having gone through superradiance.
In this type of analysis, one must properly account for the combined effect of some astrophysical process (e.g. accretion) and the black hole age.
Ref.~\cite{Ng:2020ruv} used the 45 LIGO-Virgo detections from the first part of their third observing run, to exclude the existence of ultralight scalar bosons in the mass range $[1.3-2.7]\times 10^{-13}$~eV.
In addition, by reducing the spin of individual black holes, superradiance could reduce the recoil velocity of merging binary black holes, and thus increase the retention fraction of hierarchical merger remnants, leaving an imprint on the population of black holes in dense stellar environments \cite{Payne2021}.

In the next 10-15 years, existing constraints will become significantly tighter, and discovery may be possible. The sensitivity of ground-based detectors will increase significantly with the upgrade of LIGO, Virgo, and KAGRA to their ``plus'' configurations in the next 5 years~\cite{KAGRA:2013rdx} promising hundreds of binary black hole merger measurements per year, and thus increasing the statistical power of population-based analyses. Indeed, Refs.~\cite{Arvanitaki:2016qwi,Ng:2019jsx} have shown that hundreds of high signal-to-noise ratio sources are required for a highly significant detection of a boson with mass between $\sim10^{-13}-10^{-11}$~eV. In the 2030s, next-generation detectors such as Cosmic Explorer and Einstein Telescope will yield up to $\sim 10^5$ signals per year~\cite{Ng:2020qpk}, some of which with very high signal-to-noise ratio and exquisite spin measurements~\cite{Vitale:2016icu}. By that time we will also know more about the astrophysical formation channels of binary black holes, which might reduce correlations between astrophysical uncertainties and the dark matter parameters we would like to constrain by studying superradiance.

In a different mass range, supermassive black holes can be used to probe lighter bosons.  LISA is expected to collect data from supermassive black hole binaries in the mid 2030s, and to set constraints on a wider boson mass range. Specifically, early studies indicate that black hole spin measurements with LISA will probe boson masses in the complementary range $\sim [10^{-18},10^{-14}]$ eV~\cite{Brito:2017zvb}. 
To fully benefit from this wealth of data we need a better theoretical and numerical understanding of boson clouds in black hole binaries. This will ensure that systematical uncertainties in our modeling of the emission process will not become the limiting factor. 

\subsubsection{Numerical and theoretical opportunities in superradiance}

Much progress has been made in understanding the superradiant instability of ultralight bosons around spinning black holes, including calculating the instability growth rates and 
gravitational wave signals using techniques from black hole perturbation theory, and studying the growth and saturation of the instability in the absence of non-gravitational interactions using nonlinear simulations.  However, a number of open problems remain.  A major limitation in accurately modelling the gravitational wave signals from the oscillations of boson clouds is the lack of a robust calculation of the frequency evolution which will occur as the cloud dissipates.  Currently, the change in frequency is estimated based on the Newtonian gravitational self-energy of the bosonic cloud ~\cite{Baryakhtar:2017ngi,Baryakhtar:2020gao,Siemonsen:2019ebd}, which is not accurate for the loudest signals.  This limits the amount of time over which one coherently integrates a putative signal, and hence the sensitivity of current searches.

Another forward direction is the study of how ultralight boson self-interactions or couplings to other matter may affect the superradiant instability and subsequent gravitational wave emission. Scalar or vector self-interactions modify the cloud's dynamics, possibly leading to instabilities, multi-mode configurations and premature interruption of the growth of the cloud, as well as suppressed gravitational and new scalar wave signals ~\cite{Baryakhtar:2020gao,Fukuda:2019ewf}.
It also has been suggested that a coupling between the Standard Model photon and a superradiant scalar boson with an axionic interaction~\cite{Rosa:2017ury,Sen:2018cjt,Ikeda:2018nhb,Boskovic:2018lkj,Chen:2019fsq,Chen:2021lvo}, or a kinetic mixing between the photon and a dark massive photon~\cite{Caputo:2021efm}, may lead to potentially observable electromagnetic emission and signatures in the polarimetric imaging of supermassive BHs through EHT-like observations~\cite{Chen:2019fsq,Chen:2021lvo}. However, most of the current studies have relied on simplified models or taking a non-relativistic limit to make the nonlinear calculations tractable, while nonlinear solutions of the full equations have been restricted to purely gravitational interactions and the relativistic regime~\cite{Zilhao:2015tya,East:2017ovw,East:2018glu}. There are only a few studies of simplified models showing the possible impact of spatially varying plasmas~\cite{Dima:2020rzg,Cannizzaro:2020uap,Cannizzaro:2021zbp,Wang:2022hra}, and finding better techniques to tackle the disparate timescales in the problem while capturing all relevant nonlinear and back-reaction effects (e.g.~\cite{Cardoso:2020nst,Blas:2020kaa}) is an important direction for future work. An outstanding question is in what, if any, circumstances a so-called ``bosenova" may occur, where nonlinear effects lead to a violent collapse/explosion of the ultralight boson cloud~\cite{Arvanitaki:2010sy,Yoshino:2012kn,Baryakhtar:2020gao,Omiya:2020vji,Omiya:2022mwv}.

As alluded to above, another opportunity is studying the effect of the cloud on a binary inspiral. The gravity of the companion perturbs the cloud~\cite{Baumann:2018vus}, which in turn induces a backreaction on the orbit in the form of resonant (``floating'' or ``kicked'') orbits \cite{Zhang:2018kib,Baumann:2019ztm}, as well as a continuous friction \cite{Zhang:2019eid,Baumann:2021fkf,Tahura:2022ffs} which speeds up the inspiral. Both of these effects can have a dramatic impact on the dynamics of the system, which can differ significantly from the expectations in vacuum. Moreover, finite size effects change the measurable tidal response and the quadrupole moment of the central black hole \cite{Baumann:2018vus,DeLuca:2021ite,Su:2021dwz}, and affect the mass and momentum of the companion, if it is a black hole~\cite{Baumann:2021fkf}. To leverage these promising channels of detection, however, we need a combined and consistent treatment of this wealth of interconnected phenomena to achieve a full understanding of the phenomenology. This work is a necessary preliminary to the systematic modelling of gravitational waveforms.

\subsection{Novel signals of bosonic dark matter }

Driven by gravity, ultralight bosons can form bound states---boson stars (BSs)---comparable in mass and compactness to neutron stars and black holes. Boson star binary inspirals emit gravitational waves detectable by LIGO, Virgo and KAGRA for boson masses within $10^{-9}-10^{-13}$ eV, and by space-based gravitational wave detectors like LISA for masses within $10^{-20}-10^{-14}$ eV. Beyond their gravitational wave emissions, stars made of axions coupling to the photon can leave observational electromagnetic imprints as well as send out axion bursts, when exploding or merging with neutron stars. These could be detected by axion DM detectors in the mass range $10^{-15}-10^{-7}$ eV.

\subsubsection{Boson stars}
\label{sec:bosonstars}

Boson stars are gravitationally bound clumps of condensates of ultralight scalar or vector particles \cite{Ruffini:1969qy,Kaup:1968zz,Brito:2015pxa}. Provided an abundance of ultralight bosons \cite{Hui:2016ltb,Li:2013nal}, gravitational interactions lead to the formation of condensed structures with large occupation numbers \cite{Levkov:2018kau} -- boson stars. Typical spherically symmetric configurations in linear models can attain astrophysical masses $\sim (10^{-11} \mathrm{eV}/m_b) M_\odot$, given a boson mass $m_b$. Scalar or vector self-interactions, as well as angular momentum strongly affect this relation and open up a wide range of possible configurations
 \cite{Schunck:2003kk}. For the purposes of gravitational wave observations, this large set of stars can be classified by their compactness, i.e., the mass in relation to their radius, $C=GM/Rc^2$. The latter varies from the Newtonian regime $C\sim 10^{-6}$ (comparable to our sun), to the ultra compact regime with $C\sim 0.5$ (comparable to black holes).

Across the BS parameter space, gravitational wave observations of binary BSs (making no assumptions about potential couplings of the bosonic field to the Standard Model) are ideally suited to find evidence for the existence of these objects, and therefore, also evidence for a new ultralight boson. For binary BSs with constituent compactnesses comparable or above those of neutron stars, $C> 0.1$,

gravitational waves from the merger and post-merger phases are promising for a confident detection. In those scenarios, the signals are highly dependent on the internal structure of the compact object, and hence, signatures left by binary black holes or binary neutron stars are expected to be readily distinguishable from those of BSs with the Einstein Telescope/Cosmic Explorer and LISA \cite{Toubiana:2020lzd}. 
The gravitational wave signals can probe the global structure of the non-linear potential of the dark sector through its imprints on the mass profile of the BSs~\cite{Croon:2018ybs} as well as through the stochastic gravitational wave production~\cite{Croon:2018ftb}.  

The smoking gun gravitational wave signals range from slowly decaying oscillations of a BS remnant \cite{Cardoso:2016rao,Macedo:2016wgh} to echos \cite{Cardoso:2016oxy, Urbano:2018nrs}. Despite their promise, however, only a handful of necessary numerical simulations of merging BSs and resulting gravitational waveforms exist \cite{Palenzuela:2007dm,Bezares:2017mzk,Palenzuela:2017kcg,Dietrich:2018bvi,Bezares:2018qwa,Helfer:2021brt,Liebling:2012fv}, greatly reducing detectability prospects with matched-filtering techniques. Furthermore, while the dynamical stability and formation of non-spinning stars is well-understood \cite{LEE1989477,PhysRevD.38.2376,GLEISER1989733,PhysRevLett.72.2516,Guzman:2004jw,Guzman:2006yc,Bernal:2006it,Balakrishna:1997ej}, corresponding studies for rotating stars \cite{Sanchis-Gual:2019ljs,DiGiovanni:2020frc, Purohit:2021hpg} revealed only scalar BSs with self-interactions are stable \cite{Siemonsen:2020hcg,Dmitriev:2021utv}; how rotating BSs form dynamically is still an open problem. Isolated rotating BSs may undergo an ergoregion instability \cite{Friedman:1978} and leave observational imprints on the stochastic gravitational wave background \cite{Barausse:2018vdb}. Even during the early binary inspiral, less compact BSs with $C< 0.1$ can be distinguished by their tidal interactions from black hole and neutron star binaries with current generation gravitational wave observatories, while the Einstein Telescope/Cosmic Explorer and LISA will enable to distinguish compact BSs during the inspiral \cite{Sennett:2017etc}. Additionally, these stars may undergo inspiral resonances that lead to potentially detectable variations in the gravitational wave power emitted \cite{Macedo:2013jja,Macedo:2013qea}. Finally, gravitational waves from BSs with $C< 10^{-2}$ orbiting supermassive black holes are expected to be detectable with LISA \cite{Guo:2019sns}. Once detected, the form of the non-linear potential can be probed analogous to neutron star equations of state from their mass-radius relationships \cite{Croon:2018ybs}. Microlensing allows for significant synergy in compact object searches associated with new physics, including boson and axion stars~\cite{Croon:2020wpr,Croon:2020ouk,Fujikura:2021omw}, axion miniclusters~\cite{Fairbairn:2017dmf} as well as primordial black holes~(e.g.~\cite{Niikura:2017zjd,Sugiyama:2020roc,Kusenko:2020pcg}).
 
\subsubsection{Axion astronomy with transient sources}
\label{sec:axionast}

Boson (axion) stars can generally contribute a variety of distinct signatures, allowing for new detection opportunities beyond conventional cold dark matter as well as multimessenger prospects beyond just gravitational waves. With axion-photon coupling, electromagnetic signatures can appear from resonant photon production associated with collapsing axion stars~\cite{Levkov:2020txo} or neutron star-axion star collisions~\cite{Dietrich:2018jov}. Explosions of axion stars and other transient sources can naturally lead to a new class of observables~\cite{Eby:2021ece} in terrestrial experiments associated with burst emission of relativistic axions~\cite{Eby:2016cnq, Levkov:2016rkk}.
For the QCD axion, bursts from collapsing
axion stars lead to potentially detectable signals over a wide range of axion masses $10^{-15}~{\rm eV} \lesssim m_a \lesssim 10^{-7}$~eV in future experiments, such as ABRACADABRA~\cite{Salemi:2021gck}, DMRadio and SHAFT~\cite{Gramolin:2020ict}. Unlike
conventional cold axion DM searches, the sensitivity to axion bursts is not necessarily suppressed as
$1/f$ for large decay constants $f$.
More so, unlike traditional cold dark matter searches, axion potential self-interactions sensitively affect axion burst emission signals as well as their spectra. Thus, intriguingly, the detection of axion bursts could provide new insights into the
fundamental axion potential~\cite{Eby:2021ece} that is challenging to probe otherwise.
Furthermore, in analogy with neutrino astronomy (e.g. \cite{Munoz:2021sad}), emission of relativistic axions from historic transient sources results in yet another class of signatures - the diffuse axion background - distinct from conventional cold axion dark matter~\cite{Eby:2021ece}. As experimental searches typically focus primarily on detection of conventional cold dark matter, they can miss these new types of signatures. It is thus imperative to expand the experimental analysis toolbox in the future to cover a broader range of observables.

Beyond the BSs themselves as a transient source, axions with a photon coupling could lead to signatures observable in other transient sources. 
During the propagation of photons emitted by transient sources such as supernovae, the photon flux could be either suppressed or enhanced depending on the axion mass range. 

For axions with mass $m \lesssim 10^{-13}\mathrm{eV}$, the conversion of photons to axions during the photon propagation through the magnetic field background leads to a dimming of the magnitude of sources such as Type IA SNe and galaxy clusters. This in turn alters the inferred cosmic distance and can be constrained by cosmic distance data sets~\cite{Buen-Abad:2020zbd}. This puts constraints on the axion-photon coupling to be below $G_{a\gamma\gamma} \lesssim  10^{-11}-10^{-12} \mathrm{GeV}^{-1}$. Future improvements in the understanding of the magnetic field in the Inter-Galactic Medium (IGM) and Intra-Cluster Medium (ICM) could potentially tighten these constraints. 

On the other hand, if the axion DM mass is around $10^{-6}-10^{-5}\,\mathrm{eV}$, the decay of axion dark matter stimulated by the radio photon waves (\textit{i.e.} that from ancient supernova remnants) results in radio echos that are observable at SKA Phase I and FAST radio telescope~\cite{Sun:2021oqp,Buen-Abad:2021qvj}. This results in a constraint on the coupling between axion DM and photons to be $G_{a\gamma\gamma} \lesssim  10^{-10}-10^{-11} \mathrm{GeV}^{-1}$. Better understanding of faint radio sources background and denser short baselines (such as SKA Phase II) could potentially further improve the bounds. 

Lastly, axions are expected to be copiously produced at the core of SNe. When they propagate through the galactic magnetic field, the conversion to gamma ray photons leaves imprints on the gamma ray spectrum. This in turn puts strong bounds on the coupling between axions and photon. Some of the bounds include that from SN1987A~\cite{Grifols:1996id,Brockway:1996yr,Payez:2014xsa}, and more recent constraints obtained using AGN and Chandra observations~\cite{Reynes:2021bpe}, Fermi-LAT spectral analysis of NGC 1275\cite{Fermi-LAT:2016nkz}, and search of extraglactic core-collapse SNe with Fermi-LAT~\cite{Meyer:2020vzy}. Using the gamma ray burst data, one can also constrain axions produced inside the core-collapse SNe~\cite{Meyer:2016wrm,Crnogorcevic:2021wyj}.

\subsection{Stellar evolution signatures of ultralight bosons}
\label{sec:ULDMstellar}

Stellar evolution offers powerful ways to study the physics of ultralight, weakly  interacting particles~\cite{Raffelt:1996wa,Agrawal:2021dbo}. 
In fact, stars are sensitive to extremely rare processes, which are often prohibitively difficult to be observed in colliders. 
Axions and ALPs are a notorious example. 
Only very recently were terrestrial experiments able to probe 
the axion parameter space in regions not excluded by stellar evolution~\cite{CAST:2017uph,Giannotti:2017law}, and very promising outcomes are expected in the near future~\cite{DiLuzio:2021ysg}.

In this section, we focus on signatures derived from the evolution of low mass stars. 
Furthermore, we will refer entirely to the case of axion and ALPs, although several results apply also to dark photons and other weakly interacting particles (see Ref.~\cite{Agrawal:2021dbo}).
Notice that the stellar arguments discussed in this section do not require ALPs to be a fraction of the dark matter in the universe, since they are produced directly in the stellar core. 

Among the most valuable observables to understand the properties of ALPs and other weakly coupled particles are the tip of the red giant branch (RGB) and the R-parameter, which corresponds to the number ratio of horizontal branch (HB) versus RGB stars in globular clusters.
The RGB tip has been used in the past decades to constraint the axion-electron coupling. 
The most stringent bound on this coupling, $G_{aee}\lesssim 1.5\times 10^{-13}$, was derived in two independent contributions~\cite{Straniero:2020iyi,Capozzi:2020cbu} in 2020, taking 
advantage of the new cluster distances determination from the GAIA DR2 data~\cite{chen2018}.
Concerning the axion-photon coupling, the strongest bound  $G_{a\gamma\gamma}\leq 6\times 10^{-10}\,{\rm GeV^{-1}}$, was derived in Ref.~\cite{Ayala:2014pea,Straniero:2015nvc}, using observations of the R-parameter in several globular clusters. 
Both bounds apply to axion masses up to a few 10 keV. 
At higher masses, the constraints relax, due to the Boltzmann suppression of the production rate. 
A reliable extension of the RGB bound at high masses is still missing, though a preliminary study can be found in Ref.~\cite{Carenza:2021osu}.
The HB bound on the axion-photon coupling has been extended to higher mass only very recently, in Ref.~\cite{Lucente:2022wai}.\footnote{See Refs.~\cite{Cadamuro:2011fd,Carenza:2020zil} for earlier attempts.}

An intriguing aspect of stellar evolution observations  is a series of independent observations of stellar population that have shown unexpected behaviors
explainable in all cases as an excessive energy loss.
This led to speculations about new physics in the form of an exotic weakly interacting particle which could efficiently contribute to the stellar energy loss (see Refs.~\cite{Giannotti:2017hny,Hoof:2018ieb,DiVecchia:2019ejf,DiLuzio:2020wdo,DiLuzio:2020jjp,DiLuzio:2021ysg} for recent reviews). 
The first of these anomalies was originally reported in the analysis of the period change of the white dwarf variable (WDV) G117-B15A, interpreted in terms of axions in Ref.~\cite{Isern:1992gia}. 
Other independent observations have, since then, confirmed the trend in other WDV stars~\cite{Corsico:2019nmr};
the WD luminosity function (WDLF)~\cite{Bertolami:2014wua,MillerBertolami:2014oki,Isern:2018uce,Isern:2020non};
red giant branch (RGB) stars~\cite{Straniero:2020iyi};\footnote{Notice, however, that the hint in this case has very little significance 
and has not been reported in the analysis of Ref.~\cite{Capozzi:2020cbu}.} 
horizontal branch (HB) stars~\cite{Ayala:2014pea,Straniero:2015nvc};
red clump stars~\cite{Mori:2020qqd};
helium burning intermediate mass stars~\cite{Friedland:2012hj,Carosi:2013rla}; 
and supernovae (SN) progenitors~\cite{Straniero:2019dtm} (see also discussion in Sec.~3.4.3 of Ref.~\cite{LSSTDarkMatterGroup:2019mwo}).
The axion case is especially compelling since,
contrarily to other new physics candidates, fits particularly well all the  observations~\cite{Giannotti:2015kwo}.
In this case, the combined analysis of observations from HB and RGB stars, the WDLF, and all the WDV for which the rate of the period change was measured, 
indicates a preference for a non-zero axion-electron coupling, $G_{aee}\simeq 10^{-13}$, and axion-photon coupling,   $G_{a\gamma\gamma}\simeq 2\times 10^{-11}\,{\rm GeV^{-1}}$, with a significance of roughly 3$\,\sigma$~\cite{DiLuzio:2021ysg}.
These results should be taken with caution.\footnote{Another source of uncertainty could be the presence of magnetic fields in stellar cores, which can affect the production of both neutrinos and axion-like particles~\cite{Kachelriess:1997kn,Caputo:2020quz,Drewes:2021fjx}.}
However, they do show a systematic problem with the present understanding of stellar cooling and 
demand further investigations.

Rapid advances in experimental astrophysics promise a significant improvement in our understanding of ultralight weakly interacting particles 
and it is likely that, if such particles exist and have masses below a few keV, their impacts on stars might be revealed with high significance in stellar observations. 
This fact, combined with the enormous interest in new dedicated terrestrial experiments, promises considerable improvements in our understanding of the physics of light, weakly interacting particles in the coming decade.
Data from the GAIA survey, particularly the new cluster distances available through the Gaia DR2 data, have already allowed to revise the analysis of the impact of ALPs on Red Giant stars~\cite{Capozzi:2020cbu,Straniero:2020iyi}.
These analyses have reinforced the bound on the axion-electron coupling and reduced the significance of the hint. 
A further improvement is likely to follow, in the near future, thanks to the increased angular resolution of the next-generation
space-based missions, such as JWST~\cite{JWST:Gardner:2006ky}, which will enlarge the statistical sample of
RGB members near the cores of GCs. 
An even more substantial improvement is expected in the case of the WDLF.
The current analyses of the impact of axions on the WDLF are based on old and often inconsistent data~\cite{Bertolami:2014wua}. 
The largely increased WDs catalog, with precisely measured distances available through the GAIA data, will enable a considerable improvement of the WDLF in the coming years. 
Such improvement will also gain a significant boost 
with the starting of operations of the Vera Rubin Observatory, expected to detect WDs that are
5 to 6 magnitudes fainter than those detected by Gaia and to increase the census of WDs to tens of millions~\cite{LSSTDarkMatterGroup:2019mwo}.

Dedicated axion experiments of the next generation, particularly axion helioscopes~\cite{IAXO:2019mpb,IAXO:2020wwp} will also be able to access at least sections of the parameter space relevant for stellar evolution, offering a complementary way to study the impact of axions on stars~\cite{DiLuzio:2021ysg}. Moreover, a population of ALPs or dark photons can be produced in the core and accumulate in gravitationally bound orbits around the sun (and other astrophysical bodies). Such basin can increase the prospects of both direct and indirect detection for some mass ranges~\cite{VanTilburg:2020jvl,Lasenby:2020goo}.

\section{Light Dark Matter (keV--MeV)}

\subsection{Constraining LDM through core-collapse supernovae}
\label{sec:SNe}

Core collapse supernovae, originating from stars with mass larger than $8 \,M_\odot$, are extremely efficient factories of feebly interacting particles, 
e.g. neutrinos, axions and ALPs, and dark photons~\cite{Raffelt:1999tx,Raffelt:1996wa}. At the end of its life cycle, the compact core of an evolved star becomes
unstable and collapses to nuclear density. The standard picture (see e.g.~\cite{Janka:2012wk,Mirizzi:refrefrrf,Vitagliano:2019yzm}) is that a shock wave forms, moves outward, and---rejuvenated by the neutrino flux---ejects most of the mass in the form of a SN explosion, leaving behind a cooling proto-neutron star (PNS) which eventually becomes a neutron star (NS). Within a few seconds, the
gravitational binding energy of the NS, $E_{\rm
  b}\sim3\times10^{53}~{\rm erg}$, is released in the form of neutrinos---an energy comparable to that released by all stars in the Universe within the same
period. 
This energy is mostly emitted in the form of neutrinos (99\%), as $\gamma$ and $e^\pm$ interact so strongly that they contribute little to energy transfer. As a result, only $\sim10^{51} \, \rm erg$ are emitted as kinetic energy of the expelled material, and $\sim10^{48-49}\, \rm erg$ is emitted as photons.

The standard picture of the neutrino fluxes, energies and emission time-scale was confirmed on 23~February 1987 by the neutrino burst from
SN~1987A~\cite{Hirata:1987hu, Bionta:1987qt, Alekseev:1988gp}. Beyond the Standard Model particles, produced in the dense and hot matter of the PNS, can leave many signatures which depend on the mass and interactions of such particles. For example, the existence of axions and ALPs with coupling to nucleons (e.g.~\cite{Turner:1987by,Raffelt:1996wa,Carenza:2019pxu}) and photons (e.g.~\cite{Lucente:2020whw,Caputo:2021rux}) can be constrained. Recently, the coupling to muons of heavy ALPs has been explored~\cite{Bollig:2020xdr,Croon:2020lrf,Caputo:2021rux}. Owing to the high temperature and density of the PNS, they are the only astrophysical environments where particles with mass of up to several hundreds MeV can be abundantly produced.
If ALPs exist, the PNS may lose energy fast, affecting the duration of the neutrino signal of SN~1987A. The SN cooling argument excludes the existence of ALPs, unless their couplings are so small they are not produced efficiently (free streaming-regime) or they are so strongly interacting that they cannot escape, so that no additional cooling channel exists (trapping regime)~\cite{Turner:1987by,Mayle:1987as,Burrows:1990pk}. In the free streaming regime, a simple criterion is that the new energy loss should not exceed $10^{19}\,{\rm erg} \,{\rm g}^{-1} \,{\rm s}^{-1}$, or an overall luminosity of around few times $10^{52}\,{\rm erg}\,{\rm s}^{-1}$, to be calculated at nuclear density $\rho=3\times10^{14}\,{\rm g}\,{\rm cm}^{-3}$ and $T=30~{\rm MeV}$ \cite{Raffelt:1996wa}. For a critical take on the SN 1987A cooling bounds, see \cite{Bar:2019ifz}.

New particles can also generate visible light signals in the x-ray or gamma-ray bands~\cite{Jaeckel:2017tud}. This is possible if the new degrees of freedom are allowed to decay to SM photons. Such signal can also be used to probe ALPs with couplings to fermions, as the decay happens through a fermion loop~\cite{Caputo:2021rux}.

If the decay time is short, a beyond the Standard Model (BSM) particle can decay in the mantle around the PNS, which acts as an astrophysical calorimeter. As the kinetic energy of the expelled material is small, stringent bounds can be obtained. This argument, first advanced by Falk and Schramm~\cite{Falk:1978kf}, was recently rediscovered~\cite{Sung:2019xie}, and applied to muon-philic bosons~\cite{Caputo:2021rux}. It has been shown that muon-philic scalars, a simple solution to the observed discrepancy between the measured and predicted muon magnetic moment~\cite{Chen:2017awl,Muong-2:2021ojo}, are therefore excluded by SN energetics arguments. Similarly, the ``cosmological triangle'', the only viable region in the parameter space for MeV ALPs coupling to photons~\cite{Brdar:2020dpr}, is hardly reconciled with SN physics~\cite{Caputo:2021rux}. Very recently, it has been shown that SNe with particularly low explosion energies are the most sensitive calorimeters to constrain particle depositing energy in the mantle~\cite{Caputo:2022mah}.

Dark vectors, such as dark photons with a kinetic mixing with SM photons, can be constrained with similar arguments, so SNe provide guidance for their experimental searches~\cite{Chang:2016ntp,DeRocco:2019njg}. Also dark fermions, such as heavy millicharged fermions, can be constrained by SN cooling~\cite{Davidson:2000hf,Chang:2018rso} and detection at Earth~\cite{DeRocco:2019njg,Baracchini:2020owr}. SN 1987A has been also used to put bounds on CP-even scalars mixing with the Higgs boson~\cite{Krnjaic:2015mbs,Dev:2020eam}. On the other hand, the case of sterile neutrinos (for early approaches see e.g.~\cite{Shi:1993ee,Nunokawa:1997ct}) is more complicated, since particular care should be paid to the modeling of the SN evolution in the presence of active-sterile mixing~\cite{Suliga:2019bsq,Suliga:2020vpz}.

In some cases, strong constraints can be obtained using the diffuse x-ray and gamma-ray background, an idea dating back back to an early paper by Cowsik~\cite{Cowsik:1977vz}. All collapsing stars in the visible universe, a few per second, provide the diffuse supernova background of dark particles (similar to the diffuse supernova neutrino background~\cite{Vitagliano:2019yzm}), that can later decay (see e.g.~\cite{Calore:2020tjw,Caputo:2021rux}).

The enthusiasm for SNe as laboratories for astroparticle physics has grown in recent years, thanks to both advancements in numerical simulations~\cite{Janka:2016fox,Bollig:2017lki}
and the development of new neutrino detectors, as well as gamma-ray detectors (INTEGRAL, Fermi). Proposed future detectors like e-ASTROGAM and AMEGO are promising avenues to detect potential signals.
The next nearby SN will provide a bonanza of astrophysical and particle-physics information, being a factory of particles with mass up to several hundreds MeV. It will be observed in a large number of detectors of different size, ranging from Super-Kamiokande to IceCube, although in the latter case without event-by-event recognition~\cite{Scholberg:2012id,Scholberg:2017czd,Mirizzi:refrefrrf}. Upcoming large detectors such as Hyper-Kamiokande or DUNE also provide promising detection perspectives~\cite{Scholberg:2017czd,Mirizzi:refrefrrf}. On the theoretical side, our understanding of nuclear matter in the extreme conditions of PNS and NS is still limited, and the particle production rates are still affected by theoretical uncertainty, though development has been on-going~\cite{Fore:2019wib,Carenza:2020cis}. Moreover, improvements in the simulations and self-consistent inclusion of new strongly coupled degrees of freedom would also be welcomed.
The most extreme astrophysical events like hypernovae~\cite{Moriya:2018sig}, though rare, could speak volumes about ALPs, SN explosion mechanisms, and more~\cite{Caputo:2021kcv}.

\subsection{LDM constraints from binary neutron star mergers}
%
\label{sec:BNS}

Binary neutron star mergers are a new promising environment to probe weakly interacting light particles. Immediately after the merger, these remnants reach temperatures in the $30-100$ MeV range, and densities above $10^{14} \text{g/cm}^3$, similar to the proto-neutron stars formed in core-collapse supernovae which have been used to place constraints on a wide range of scenarios (see Sec.~\ref{sec:SNe}). The large temperature and density of these objects makes them very efficient at producing feebly interacting dark sector particles, which can escape this environment and lead to observational signals~\cite{Dietrich:2019shr,Harris:2020qim,Diamond:2021ekg}. Two key distinctions between BNS and SN are that the former allows the use of the associated gravitational wave signal as a trigger and a timing measurement to help distinguish signal from background fluctuations, and the environment around the remnant is much less baryon rich, which leads to a cleaner signal from decaying dark sector particles.

In a recent study~\cite{Diamond:2021ekg}, it was shown that dark photons with masses in the $1-100$ MeV range would be copiously produced and that for a large range of unconstrained couplings they would lead to a very bright transient gamma-ray signal originating from the dark photon decay. This new signature can be used to test most of the unconstrained parameter space in which the dark matter abundance is obtained through the freeze-in mechanism via dark photon mediated interactions, for dark photon masses smaller than $100$ MeV. There are two distinct regimes for the signal, depending on whether the decay products thermalize or not after they are produced. When the decay products thermalize, which correspond to larger couplings, most of the energy gets converted to photons following a thermal spectrum with apparent temperature around 100 keV. The parameter space for which this happens should already be probed by current instruments, such as the Gamma-ray Burst Monitor (GBM) in the Fermi satellite, once new BNS mergers are observed. The reach for the lower coupling scenario, when the decay products do not thermalize, has not been worked out in detail yet, and presents theoretical and experimental challenges. From the theory side, one needs to compute the fraction of the energy that gets converted to photons, and the spectrum of those photons. The experimental challenges stems from the fact that in this scenario most of the energy will remain in the leptons, and so the expected photon luminosity will be lower, and should also be more concentrated at larger energies, comparable to the remnant's temperature.

Future improvements in gravitational wave detectors will allow for early warning and localization of neutron star mergers, potentially allowing for instruments with narrower fields of view to observe the event. New proposed gamma-ray telescopes , such as e-ASTROGRAM~\cite{e-ASTROGAM:2016bph}, AMEGO~\cite{AMEGO:2019gny} and GECCO~\cite{Orlando:2021get} will improve the sensitivity to transient gamma-ray signals over a wide range of energies by at least an order of magnitude. In addition, they will have better photon angular resolution, which will help reduce background from diffuse gamma-rays. Both developments will make neutron star mergers an exciting target to search for light dark sectors in the upcoming decade.

\subsection{Constraining LDM using black hole population statistics}
\label{sec:BHPS}


The mass function of astrophysical black holes measured through LIGO/Virgo/KAGRA gravitational wave observations of binary black hole mergers \cite{LIGOScientific:2021psn} can be a complimentary LDM due to its effects on BH formation from massive stars \cite{Croon:2020ehi,Croon:2020oga,Baxter:2021swn,Ziegler:2020klg,Ellis:2021ztw}. Some BH progenitors reach core temperatures and densities where they experience the \textit{Pair-Instability} \cite{1967ApJ...148..803R}. Thermal production of electron-positron pairs softens the equation of state (EOS) resulting in a gravitational contraction that raises the core temperature and ignites oxygen explosively. The explosion either results in a series of mass-shedding pulsations --- a \textit{Pulsational Pair-Instability Supernova} (PPISN) --- that leaves less bound mass to form the final BH, or, in heavier objects, unbinds the star entirely leaving no BH remnant --- a \textit{Pair-Instability Supernova} (PISN). 
The absence of BHs with masses $M\gtrsim 50 {\rm M}_\odot$ is referred to as the \textit{Upper Black Hole Mass Gap} (UBHMG). Future GW data can be used to determine its existence and location \cite{Fishbach:2017zga,Fishbach:2019bbm,Baxter:2021swn,LIGOScientific:2020kqk,LIGOScientific:2021psn}.

Weakly coupled LDM particles e.g.~axions and dark photons can be produced in the cores of massive stars. They subsequently free-stream, acting as a novel source of energy loss that is compensated by an increased rate of nuclear burning. This shortens the lifetime of core helium burning, leaving less time for the (subdominant) $^{12}{\rm C}(\alpha,\gamma)^{16}{\rm O}$ reaction to proceed resulting in less $^{16}{\rm O}$ present at the onset of the pair-instability and consequentially less violent explosions. Heavier BHs can be formed as a result, and BH mass as a function of initial mass becomes more sharply peaked \cite{Croon:2020ehi,Croon:2020oga}.~In some LDM models e.g.~dark photons, the pulsations may be absent completely. 
Heavy DM ($M_{\rm DM}\gtrsim 10$ keV) can also alter the physics of PPISN/PISN either through its contribution to the EOS \cite{Croon:2020oga}, gravitational capture \cite{Ellis:2021ztw}, or via energy injected from annihilations \cite{Ziegler:2020klg}. 
LDM with significant interactions with stellar matter, such as axions in the cosmological triangle, can affect the EOS in such a way that a new instability exacerbates pair-instability. This can significantly alter late stellar evolution and ultimately both the location of the UBHMG and the luminosity of resulting supernovae \cite{Sakstein:2022tby}.

Predictions for the BHMF can be made using the stellar structure code MESA \cite{Paxton:2017eie}, suitably modified to include LDM effects \cite{Croon:2020ehi,Croon:2020oga,Sakstein:2020axg,Sakstein:2022tby}.~Constraints on the model parameters may be extracted from GW observations of the black hole mass function (BHMF) using a novel three-parameter fitting function developed by \cite{Baxter:2021swn} and comparing with MESA predictions.
To set competitive constraints, precise determination of the BHMF is needed. 
~Information from the shape of the BHMF and the location of the UBHMG can be used to mitigate stellar modelling uncertainties e.g.~metallicity and nuclear reaction rates, and can distinguish between competing LDM models. Anticipated LIGO/Virgo/KAGRA data releases will enable the first constraints to be placed using this technique.

\subsection{Heating and gamma rays from astrophysical objects through LDM scattering and annihilation}
\label{sec:heatinggamma}


DM can be captured in astrophysical objects, by scattering and losing sufficient energy to be gravitationally bound. Once captured, depending on the particle model, DM may annihilate to SM products. If the annihilation is sufficiently prompt, the annihilation products are absorbed by the astrophysical object, increasing its temperature, which can be detectable. It has been pointed out that exoplanets~\cite{Leane:2020wob}, brown dwarfs~\cite{Leane:2020wob}, population III stars~\cite{Freese:2008hb, Taoso:2008kw, Ilie:2020iup, Ilie:2020nzp}, white dwarfs and neutron stars~\cite{Goldman:1989nd,Bertone:2007ae,NSvIR:Baryakhtar:DKHNS,Bell:2021fye} can all be excellent dark matter heating detectors in complementary scenarios. Assuming the scattering process, which is necessary for capture, is in equilibrium with the annihilation process, which is necessary for sufficient heating in all objects except neutron stars (see below), the annihilation heat provides a new probe of the DM scattering rate. 

Each of these objects can probe sub-GeV dark matter scattering, though the scenario in which they are most optimal DM detectors varies. Neutron stars are incredibly dense, and so generally offer the best cross-section sensitivity reach. However, the DM-heating luminosities of neutron stars can be very low, as they are tiny objects. Their DM heat therefore cannot be as easily detected, especially at large distances. On the other hand, e.g. Jupiter-like exoplanets are about 1,000 times larger in radius, and so their DM-heating signal can be detected far into the Galactic center. As the amount of DM available directly correlates with the DM-heated temperatures of the exoplanets, this allows for a new probe of the DM density distribution throughout the Galaxy.  
This exoplanet search is unique in its ability to potentially provide the first non-gravitational probe of the DM density distribution in our Galaxy. 
This DM-heating signal may also be detectable at upcoming infrared and optical telescopes, JWST, Roman and Rubin~\cite{Leane:2020wob}.

Another large object whose DM-heating can potentially be detected at large distances, is a white dwarf. However, they have relatively high background temperatures, such that a large amount of DM is required to heat them above backgrounds. For this reason, they are often studied in a globular cluster called Messier 4, which could potentially have large DM content, though this is not known yet and has considerable astrophysical uncertainty. Limits based on the assumption that there is a large amount of DM in Messier 4 were set in e.g. Refs.~\cite{Bertone:2007ae,Bell:2021fye,Cermeno:2018qgu}. The robustness of these bounds can be improved in future with more accurate determination of the DM content in globular clusters and reduced systematic uncertainties.

In addition to the DM heating signature discussed above, LDM can also be probed using gamma rays. In the case that the DM annihilation products are produced via decay of a sufficiently long-lived or boosted mediator, the annihilation products can be detected directly. The strong constraining power has been pointed out for sub-GeV DM scattering using Fermi-LAT data of Jupiter~\cite{Leane:2021tjj}, as well as Galactic center gamma-ray data for a Galactic population of brown dwarfs~\cite{Leane:2021ihh} (see also solar gamma-ray constraints~\cite{Leane:2017vag,HAWC:2018szf,Nisa:2019mpb}). Going forward, better determination of the Galactic distribution of brown dwarfs and neutron stars, as well as more accurate understanding of the DM density profile, is required to improve the accuracy of the inner Galaxy gamma-ray searches.

Lastly, DM annihilation to neutrinos can be observed within some of these objects, as neutrinos are sufficiently weakly interacting to escape the object's cores. In the case that DM annihilates to long-lived particles, more energetic neutrinos can escape which would otherwise be heavily attenuated~\cite{Bell:2011sn,Leane:2017vag}. Improvements to the existing sensitivity to neutrinos with Super-K~\cite{Super-Kamiokande:2015xms}, IceCube~\cite{IceCube:2016dgk}, and ANTARES~\cite{ANTARES:2016xuh}, can be achieved with KM3Net~\cite{Leane:2017vag}, and as well as other upgraded neutrino detectors in future.

\subsection{Heating of neutron stars through Auger effect}
\label{subsec:AugerNS}


Dark sectors containing GeV-mass states carrying baryon number could explain the long-standing neutron lifetime anomaly~\cite{Fornal:2018eol} and 
the recent XENON1T excess~\cite{McKeen:2020vpf},
and feature in solutions to baryogenesis~(see, e.g., \cite{McKeen:2015cuz,Alonso-Alvarez:2021oaj}).
One consequential species is the dark neutron, which mixes with the standard neutron and could arise as an elementary particle~\cite{Fornal:2018eol} or as a composite in mirror world scenarios~\cite{Berezhiani:2018eds}.
For tiny mixing, the dark neutron is sufficiently long-lived to constitute the dark matter of the universe.

Dark neutrons $\chi$ may be produced in neutron stars in neutron-nucleon scattering processes $n N \to \chi N$ and neutron decay $n \to \chi \gamma$.
If these proceed at high rates, the presence of the $\chi$ fluid in chemical equilibrium with nucleons would generically soften the equation of state of NS matter, which is incompatible with observations of high-mass NSs~\cite{McKeenNelsonReddyZhouNS,SheltonNS,MottaNS,EllisPattavinaNS,Berryman:2022zic}.
But if these processes occur on timescales longer than NS lifetimes (typically Myr-Gyr), overlapping with parametric regions where they also constitute DM, they would leave a visible astrophysical signature: NS overheating.
As nucleons leave behind holes in their Fermi sea, either via conversion to $\chi$ or upscattering, surrounding nucleons of higher energy rush to fill them in, releasing heat in the form of electromagnetic and kinetic energy.
This ``nucleon Auger effect" could prevent NSs from cooling down, and Hubble Space Telescope observations of the coldest (40,000 K) pulsar PSR J2144-3933~\cite{coldestNSHST} already place strong constraints on dark neutrons~\cite{McKeen:2020oyr,McKeen:2021jbh}.
These sensitivities could be improved with current and upcoming telescopes in the ultraviolet, optical and infrared that are suited to measure colder NSs: LUVOIR~\cite{theluvoirteam2019luvoir}, Rubin~\cite{Rubin1,Rubin2}, DES~\cite{DES:2019rtl}, Roman~\cite{green2012widefield}, JWST~\cite{JWST:Gardner:2006ky}, TMT~\cite{TMT:2015pvw}, and ELT~\cite{ELT:neichel2018overview}.
Dark neutrons, along with DM capture described in Sec.~\ref{subsec:captureNS}, provide compelling fundamental physics motivations for these campaigns to undertake systematic measurements of NS luminosities. 

\subsection{LDM-driven collapse of astrophysical objects}
\label{sec:LDMcollapse}

A number of studies have identified sub-GeV mass asymmetric DM models that cause neutron stars and other astrophysical objects to collapse through accumulation of dark matter, followed by the internal collapse of the captured DM, and the resulting formation of a small black hole which grows to consume the astrophysical object/NS \cite{Goldman:1989nd,Gould:1989gw,Bertone:2007ae,deLavallaz:2010wp,Kouvaris:2010jy,Kouvaris:2011fi,Kouvaris:2011gb,McDermott:2011jp,Bramante:2013hn,Bell:2013xk,Bramante:2013nma,Bramante:2015dfa,NSvIR:GaraniGenoliniHambye,Acevedo:2019gre,Janish:2019nkk,Acevedo:2020gro}. The relevant dark matter dynamics, associated signatures, and potential astrophysical searches are detailed in Section \ref{sec:dmicocollapse}. Here we review considerations specific to LDM that collapses astrophysical objects. The LDM models which have been identified as leading to the collapse of accumulated DM in NS interiors include bosonic dark matter or fermionic dark matter with substantial attractive self-interactions. Fermions would require some additional attractive force, because sub-GeV mass fermions without attractive self-interactions would require the accumulation of more than a solar mass of DM to initiate collapse, since fermions are stabilized by Fermi degeneracy pressure. In the case of light bosonic DM, DM self-interactions can substantially alter the amount of DM required for collapse inside a NS, since these can result in a substantially repulsive force between the bosons \cite{Bramante:2013hn,Bell:2013xk,Garani:2018kkd}. For the case of light fermionic DM with attractive self-interactions, it has been shown that, for certain Yukawa interactions mediated by a light scalar, the formation of a small black hole is inhibited during collapse by relativistic effects \cite{Gresham:2018rqo}. For most LDM parameter space, the signatures associated with $e.g.$ neutron star collapse, are similar to signatures from heavier dark matter discussed at length in Section \ref{sec:dmicocollapse}.

\section{Heavy Dark Matter (\texorpdfstring{$\gtrsim$ GeV}{})}

\subsection{Capture in neutron stars: heating signatures}
\label{subsec:captureNS}


\begin{table*}[t]
    \centering
    \begin{tabular}{|c|c|c|c|}
\hline
        effect & change in capture rate & applicability & reference \\
         \hline
         \hline
       \multirow{2}{*}{EoS of star effects} & $\mathcal{O}$(1): BSk20 $\to$ 21 & all $m_\chi$ & \cite{NSvIR:GaraniGenoliniHambye} \\
         & none: QMC-2 $\to$ BSk24 & all $m_\chi$ & \cite{NSvIR:anzuiniBell2021improved} \\
        \hline
        mass-radius configuration   & $\mathcal{O}(100)$ as $1 \to 2.2 M_\odot$ & all $m_\chi$ & \cite{NSvIR:Bell:Improved} \\
        \hline
       nuclear self-energies  &\multirow{2}{*}{30$-$100} & $m_\chi >$~100 MeV, any EoS &\cite{NSvIR:Bell2020improved}  \\
    nucleon structure & & $\mathcal{O}(10^3)$ for 2 $M_\odot$ NSs & \cite{NSvIR:anzuiniBell2021improved} \\
        \hline
        non-elastic scattering & subdominant & $-$ & \cite{NSvIR:anzuiniBell2021improved}\\
        \hline
        ``collective" effects & $\mathcal{O}(1-10^3)$ & 2 $M_\odot$ NS, & \cite{DeRocco:2022rze} \\
        &  &  $m_\chi < 100$~MeV, & \\
         & &  $A'$ mediator & \\
        \hline
        superfluidity: energy gap & maybe $\mathcal{O}$(1) & $m_\chi \lesssim 35$~MeV, & \cite{NSvIR:Pasta} \\
        & & single phonon excitation & \cite{NSvIR:clumps2021} \\
        \hline
       NS opacity/ extinction factor & $\mathcal{O}(1)$ & $m_\chi >$~GeV & \cite{NSvIR:Bell:Improved} \\
        \hline
       \multirow{2}{*}{relativistic kinematics} &  $\sim 4$ & $m_\chi >$~GeV & \cite{NSvIR:Bell:Improved} \\
       &  $\sim 10$ & $m_\chi <$~GeV & \cite{NSvIR:Bell:Improved}\\
        \hline
      gravitational focusing  & $< 2$ & all $m_\chi$& \cite{NSvIR:Bell:Improved} \\
        \hline
    \multirow{2}{*}{light mediator kinematics} &  $\mathcal{O}(1)$ & $m_\phi/\mu_{\rm red} < 10^{-1}$ &  \multirow{2}{*}{\cite{NSvIR:DasguptaGuptaRay:LightMed}} \\
    & voided & $m_\phi/m_\chi < 10^{-4}$ & \\
        \hline
    \end{tabular}
    \caption{A non-exhaustive list of effects that modify the rates of dark matter capture in neutron stars via nucleon scattering.}
    \label{tab:NSsideuncerts}
\end{table*}

By virtue of their extreme densities, steep gravitational potentials, and typically cold temperatures, neutron stars are excellent captors and thermal detectors of particle dark matter.
The capture of DM in NSs and its subsequent thermal relaxation was first treated in Ref.~\cite{Goldman:1989nd}.
It was recently realized that a simple probe of dark matter scattering on Standard Model (SM) states is the heat generated in the NS from the transfer of DM kinetic energy to the NS's constituent particles during the infall of DM at semi-relativistic speeds~\cite{NSvIR:Baryakhtar:DKHNS}.
It was also realized that upcoming infrared telescopes, 
e.g., the James Webb Space Telescope (JWST)~\cite{JWST:Gardner:2006ky}, 
the Thirty Meter Telescope (TMT)~\cite{TMT:2015pvw}, 
and the Extremely Large Telescope (ELT)~\cite{ELT:neichel2018overview} 
are sensitive to this ``dark kinetic heating" mechanism~\cite{NSvIR:Baryakhtar:DKHNS}; a dedicated sensitivity study at the recently launched JWST can be found in Ref.~\cite{NSvIR:IISc2022}.
These observations could be made following the discovery of old, isolated neutron stars in radio telescopes such as FAST~\cite{FAST2011} and CHIME~\cite{CHIME2021}.

Furthermore, in cosmological scenarios where DM collects predominantly in subhalos, for instance when the small-scale power is enhanced by an era of early matter domination, encounters between subhalos and NSs could brighten a fraction of Galactic NSs to optical and ultraviolet luminosities. 
This would be observable in all-sky surveys by current and imminent telescope missions such as 
Dark Energy Survey~\cite{DES:2019rtl},
Rubin~\cite{Rubin1,Rubin2},
and LUVOIR~\cite{theluvoirteam2019luvoir}.
Further to NS heating by dark neutrons via the Auger effect outlined in Sec.~\ref{subsec:AugerNS}, kinetic heating by DM capture gives another important motivation for these collaborations to perform precision measurements of NS luminosities.

Much effort has been concentrated in recent years on the particle physics implications of this probe.
It has been shown that orders-of-magnitude improvement over terrestrial direct detection searches may be achieved for DM with scattering that is spin-dependent and/or velocity-dependent~\cite{NSvIR:Raj:DKHNSOps}, 
inelastic~\cite{NSvIR:Baryakhtar:DKHNS,NSvIR:Bell2018:Inelastic}, 
and on electrons~\cite{NSvIR:Bell2019:Leptophilic,NSvIR:Riverside:LeptophilicShort,NSvIR:Riverside:Leptophiliclong,NSvIR:Bell:ImprovedLepton}.
Additionally, DM with muon-philic interactions~\cite{NSvIR:GaraniHeeck:Muophilic} and self-interactions~\cite{NSvIR:SelfIntDM} may be extensively probed.


The sensitivity of neutron star heating to elastic dark matter-nucleon scattering greatly complements direct detection searches in both the light and heavy limits, and could probe regions below both the spin-independent and spin-dependent xenon neutrino floors arising from irreducible neutrino backgrounds these experiments would run into in the near-future.
These features are qualitatively preserved even if the NS core, whose exact composition and phase transition history is unknown due to uncertainties in the state of QCD matter at extreme densities, does not allow DM capture due to suppressed scattering interactions.
In that case the NS crust, a much more robustly understood stellar region, still serves as a sensitive thermal detector with DM capturing via quasi-elastic nucleon scattering in the pasta layer and via superfluid phonon excitations in the inner crust~\cite{NSvIR:Pasta}.
Further studies related to the nuclear astrophysics of candidate NSs may be found in Refs.~\cite{NSvIR:Hamaguchi:Rotochemical,NSvIR:GaraniGenoliniHambye,NSvIR:Queiroz:Spectroscopy,NSvIR:Bell:Improved,NSvIR:Bell2020improved,NSvIR:anzuiniBell2021improved,DeRocco:2022rze}.
In Table~\ref{tab:NSsideuncerts} we collect various effects mentioned or explored in the literature that may modify the cross section sensitivities; much of these is ripe for future theoretical investigation.

Another important particle physics aspect is the possibility of dark matter annihilations~\cite{Kouvaris:2007ay,deLavallaz:2010wp}.
If the captured DM comes from a symmetric population, it could self-annihilate to SM final states within the NS and raise its luminosity, thereby saving telescope integration times by up to an order of magnitude~\cite{NSvIR:Baryakhtar:DKHNS,NSvIR:Raj:DKHNSOps}.
This is however contingent on the captured DM sinking down to a small thermal volume within the NS lifetime and annihilating efficiently; the thermalization of DM inside the NS is a non-trivial and model-dependent process~\cite{Bertoni:2013bsa,NSvIR:GaraniGuptaRaj:Thermalizn}.
It has been noted that a thermal Higgsino of 1.1 TeV mass, a true electroweak WIMP that has survived all constraints (see, e.g., Ref.~\cite{Krall:2017xij}), would thermalize with the NS crust quickly enough and thus show up in annihilation-induced NS heating~\cite{NSvIR:Pasta}.
Other models explored in this context are given in Refs.~\cite{Cermeno:2017ejm,NSvIR:Marfatia:DarkBaryon,NSvIR:DasguptaGuptaRay:LightMed,NSvIR:zeng2021PNGBDM,NSvIR:Queiroz:BosonDM}.

\subsection{Signatures of DM spikes around black holes}
\label{sec:DMspikes}


Depending on how they are formed, black holes may be surrounded by spikes of particle DM, which can reach enormous densities far exceeding those typically found in smooth galactic halos. The annihilation rate of DM particles scales with the square of the number density, meaning that such high densities would significantly enhance any electromagnetic signatures of annihilation. Alternatively, if such a `dressed' BH forms a binary with another compact object (such as a BH or NS), the dynamics of the binary will be influenced by the spike, or `dark dress'. Tracing out the binary dynamics through gravitational wave observations should therefore allow us to detect the presence of such dark dresses and probe their properties in such an extreme environment almost independently of whatever particle interactions the DM may have, assuming their interactions do not destroy the spike~\cite{Bertone:2019irm}.
Future GW observatories such as Einstein Telescope~\cite{Maggiore:2019uih}, Cosmic Explorer~\cite{Evans:2021gyd}, LISA~\cite{LISA:2017pwj} and TianQin~\cite{TianQin:2020hid} therefore have the ability to probe a range of particle-like and compact object DM candidates with masses heavier than roughly $m_a\sim 10^{-6}\,\mathrm{eV}$.

The mechanisms for DM spike formation depend on the nature of the BH. For astrophysical BHs, the most feasible scenario occurs through the adiabatic growth of a small BH seed at the centre of a DM halo ~\cite{1995ApJ...440..554Q,Gondolo:1999ef} (as may happen for direct collapse or Population III black holes~\cite{Bertone:2005xz}).  In this case, intermediate mass BHs ($M_\mathrm{IMBH} = 10^3 - 10^6 \,M_\odot$) are believed to be the most promising candidates for observing a DM spike today, as the baryon- and stellar-rich environments around SMBHs are expected to disrupt the spike~\cite{Ullio:2001fb,Bertone:2005hw}.  If the astrophysical conditions are such that DM spikes do persist around some SMBHs, and if we detect an extreme mass ratio inspiral (EMRI)  signal in the near future, then one can constrain various well motivated DM models like ultralight bosons, sterile neutrinos, annihilating DM, and primordial black holes  (PBHs)\,\cite{Hannuksela:2019vip, Li:2021pxf}.  For \textit{primordial} black holes (PBHs)~\cite{Green:2020jor}, the formation of a DM spike is inevitable: as the early Universe benal influence of the PBH grows. By matter-radiation equality, the DM spike will have a mass comparable to the PBH itself, and will survive unless dynamically disrupted by other black holes until lower redshifts~\cite{1985ApJS...58...39B,Mack:2006gz,Ricotti:2007jk,Boudaud:2021irr}. DM spikes are expected to have a power-law density profile down to radii close to the inner-most stable circular orbit (ISCO), with the exact normalisation and power-law slope $\gamma$ (typically $\gamma \sim 2.25 - 2.5$) depending on whether the BH is astrophysical or primordial, and on the environment in which the spike formed~\cite{Gondolo:1999ef,Adamek:2019gns}. The shape of the DM spike may therefore provide important clues to its formation mechanism.
    
Annihilation rates of interacting particle dark matter as well as resulting radiation can be significantly enhanced in the vicinity of DM density spikes around supermassive black holes (e.g.~\cite{Merritt:2002vj}) and intermediate-mass black holes (e.g.~\cite{2009NJPh...11j5016B}). Indeed, the prediction that dense DM spikes should form around PBHs has led to the conclusion that PBHs and WIMP DM are incompatible~\cite{Lacki:2010zf,Bertone:2019vsk}, since WIMP DM spikes around PBHs would give rise to huge gamma-ray fluxes which are ruled out by observations.  Assuming that a DM spike survives around the supermassive black hole M87*, ref.\,\cite{Yuan:2021mzi} used the Event Horizon Telescope observations to constrain DM annihilation.  Ref.\,\cite{Lacroix:2018zmg} studied the motion of the S2 star around the Galactic Center to constrain the parameters of the DM spike around Sgr A*.  Ref.\,\cite{Nampalliwar:2021tyz} argues that it is unlikely that a near future observation by an EHT-like array can detect the signature of a DM spike around Sgr A* via the measurement of the BH shadow radius.

Spikes of particle DM around individual PBHs may have a minor impact on the overall PBH-PBH merger rate~\cite{Kavanagh:2018ggo}, which could be detectable through GW observations. If they survive, the larger DM spikes surrounding supermassive black holes could impact the merger rates of PBH binaries embedded within them. Depending on competing effects and spike profiles, PBH-PBH merger rates could be modified by orders of magnitude compared to the galactic halo merger rates~\cite{Nishikawa:2017chy}. On the other hand, the merger rates of PBH-NS binaries embedded in a spike are negligibly affected due to weaker dependence on DM density and that DM spike spans a very limited volume~\cite{Sasaki:2021iuc}. 
    
Moreover, the gravitational waveform of a binary black hole system embedded in a dark matter spike looks different to that of a system inspiralling in vacuum. The dynamics of the system are influenced by the changing enclosed mass of the DM spike during the inspiral; by possible DM accretion onto the compact object; and by dynamical friction~\cite{Macedo:2013qea,Barausse:2014pra,Barausse:2014tra}. This latter case -- in which DM particles form a wake behind the companion compact object (black hole or neutron star) driving through the cloud -- is typically expected to dominate over the other two~\cite{Cardoso:2019rou}.
This dynamical friction drag force slows down the orbital velocity of the companion, which causes it to drop into a lower orbit earlier than it would have in vacuum. This accumulates as a `dephasing' in the gravitational waveform with respect to the vacuum signal, or equivalently as a difference in the number of cycles between a given entry frequency and merger. This dephasing effect was first studied in detail in Refs.~\cite{Eda:2013gg,Eda:2014kra}, which predicted an $\mathcal{O}(1)$ change in the number of GW cycles for intermediate mass-ratio inspirals (IMRI) systems observable with LISA. 
    
However, the motion of the companion object will inject energy into the DM spike perturbing it. These feedback processes have been modelled with, for example, the \texttt{HaloFeedback} code \cite{Kavanagh:2020cfn}. This more realistic modelling decreases the amount of dephasing with respect to the case of a static spike that remains undisturbed, leading to a percent-level effect (rather than an $\mathcal{O}(1)$ effect) on the number of GW cycles. In order to detect this dephasing, then, order millions of cycles of the systems need to be observed, so that the difference in the number of cycles from the vacuum case is non-negligible.

The GW frequency corresponding to the ISCO of an intermediate mass ratio binary consisting of an IMBH with mass $1000\,{\rm M_\odot}$ and a $1.4\,{\rm M_\odot}$ companion is approximately $4\,{\rm Hz}$.
In the case of LISA, with around 5 years of data, observing such a system between approximately $10^{-2}\,{\rm Hz}$ and ISCO will enable us to confidently distinguish the system as inspiralling through a dark matter spike as opposed to vacuum, and even enable us to characterise the density profile of the spike \cite{Coogan:2021uqv}. 

The potential of \emph{terrestrial} detectors to search for DM spikes has so far not been explored. For example, while most of the inspiral of a binary with masses $(1000, 1.4)\, \mathrm{M}_\odot$ would take place in the LISA band, the merger would occur in the band of Einstein Telescope and Cosmic Explorer, opening up the possibility of \emph{multiband searches}~\cite{Jani:2019ffg}. Such searches would combine LISA's sensitivity to the early part of the inspiral (where dynamical friction has a large impact) with a terrestrial detector's ability to precisely measure the late part of the merger (where dynamical friction matters less). This strategy has the potential to substantially improve prospects for discovering and measuring DM spikes \cite{Coogan:2021uqv}.
    
Certain PBH formation scenarios could produce black hole binaries with solar or even subsolar component masses with an intermediate mass ratio. As discussed above, PBHs are guaranteed to have DM spikes. Such a system with component masses near $(1, 0.001)\, \mathrm{M}_\odot$ would predominantly emit GWs in the frequency band of terrestrial detectors, suggesting it could be possible to measure their DM spikes with terrestrial detectors alone.
    
The observation of a spike through GW dephasing would provide important clues to the nature of the DM particles. A number of classes of DM cannot form sufficiently dense spikes to be detected and so would be immediately ruled out by a confirmed spike detection. These include ultralight bosonic `fuzzy' DM, keV-scale degenerate fermions, self-annihilating DM, and DM which is itself formed of light compact object~\cite{Hannuksela:2019vip}. Instead, the non-observation of a spike would be harder to interpret. It may be that the system under observation did not form under the right conditions to form a spike, or that the spike was formed and later destroyed by dynamical effects. It may still be possible to draw conclusions on the nature of DM from a large number of binary systems in the absence of DM dephasing; however, this would require a more detailed understanding of the prevalence of DM-dressed binaries in our Universe.
    
Initial investigations have demonstrated that DM spikes should be detectable with future GW detectors and should provide us with crucial clues about the nature of DM if they are detected. This motivates further work on the formalism for modelling DM-induced dephasing in IMRI systems. A number of refinements are required before this formalism can be reliably used to generate accurate GW waveforms:
    \begin{itemize}
        \item Analyses in the literature typically assume Newtonian dynamics for the binary. The influence of DM needs to be coupled with a relativistic description of the evolution of the binary, which in itself remains a challenge (e.g.~\cite{vandeMeent:2020xgc}). 
        \item Signals from non-circular orbits need to be calculated self-consistently. While non-zero eccentricities have been considered by a number of authors (e.g.~\cite{Yue:2019ozq,Becker:2021ivq}), the impact of eccentric orbits on the feedback on the DM spike has not yet been implemented. 
        \item  It will be necessary to understand the dephasing effects which are produced by other environmental factors~\cite{Barausse:2014tra}, such as baryonic accretion disks. With this, we can understand whether a DM overdensity can be reliably distinguished from other effects.
        \item GWs templates for searches with future detectors such as LISA will need to be accurate at the level of just a few cycles over the full inspiral. Thus, all of the above effects (even those which are subdominant) will need to be incorporated into a full self-consistent description. This will include not only dynamical friction, but also accretion and the changing mass in the portion of the DM spike enclosed by the binary's orbit.
        \item Once such a formalism has been developed, it will be necessary to \textit{accelerate} the calculations. Since the waveforms for binaries with DM spikes can be substantially dephased relative to GR-in-vacuum ones~\cite{Coogan:2021uqv}, searching for them will likely require searches distinct from those designed for black hole binaries in vacuum. While LISA will not use the same type of matched filtering searches as existing GW detectors, rapid generation of a large number of waveforms will be required to create a dedicated DM spike search pipeline. Possible approaches include surrogate models for approximating the evolution of the system.
    \end{itemize}

\subsection{Gravitational wave signatures of dark sectors in compact object mergers}
\label{sec:mergers}

The extreme environments of compact object mergers present unique opportunities to probe dark sector species, and many scenarios can be probed via their effects on the gravitational waves produced in these events. The population distributions and waveforms of gravitational radiation from mergers are sensitive to any new physics that modifies the dynamics of the merger process, encompassing a wide variety of dark sector physics.  Any GW detector that is sensitive to compact binary mergers can be used to probe this physics.  The mass range that can be probed by these types of searches depend on the particle type: for axions, the sensitive mass range is $\lesssim$ $\mathcal{O}$(eV). For other dark sector particles, the mass range probed is $\lesssim$ 10$^{-10}$ eV.

One of the most direct gravitational wave signatures arises from beyond--Standard-Model long-range forces between merging compact objects. Near the end of the inspiral process, the evolution of a compact-object binary is driven by the emission of gravitational radiation. Thus, non-gravitational contributions to the potential or to energy loss of the binary can give rise to large corrections to the gravitational-wave spectrum. This has been studied in the context of NS--NS mergers by Refs.~\cite{Hook:2017psm,Croon:2017zcu,Kopp:2018jom,Choi:2018axi}. Alternatively, similar effects can arise from modified gravity \cite{Sagunski:2017nzb,Alexander:2018qzg}.

Compact object binaries probe fifth forces roughly on the length scale of the binary separation. Thus, different classes of binary systems enable complementary constraints at different scales. The waveforms of NS--NS and NS--BH mergers are already constrained by LIGO observations of transients \cite{LIGOScientific:2021djp}, and probe fifth forces between \SI{e1} and \SI{e5} km. Many more such systems will be observed with future gravitational-wave facilities, including Einstein Telescope~\cite{Maggiore:2019uih}, Cosmic Explorer~\cite{Evans:2021gyd} and BBO \cite{Crowder:2005nr}, which will enable tight constraints on fifth forces in this regime. A similar search can be performed with measurements of the stochastic gravitational-wave background \cite{Dror:2021wrl}, which may enable comparable probes of fifth forces with longer effective ranges using massive binaries with data from LISA \cite{amaro2017laser} or from pulsar timing arrays \cite{NANOGrav:2020bcs, Shannon:2015ect, Lentati:2015qwp}.  NS--NS and NS--BH binary mergers are also sensitive to long-range forces which couple to the muon that is present inside a NS\,\cite{Dror:2019uea, KumarPoddar:2019ceq}.

Another class of signals arises from the presence of captured DM in and around neutron stars. Such DM components can constitute a significant fraction of the neutron star's total mass, and thus make a substantial contribution to the gravitational waves produced in a merger, with a distinctive spectrum and time dependence \cite{Ellis:2017jgp,Bezares:2019jcb,Karkevandi:2021ygv}. More generally, the presence of DM or other new physics effects can manifest as a modification to the equation of state of the neutron star interior, with implications for e.g., the mass--radius relation and the effective tidal deformability. These observables are sensitive to the presence of a captured DM core \cite{EllisPattavinaNS,Nelson:2018xtr} or possibly even to new contributions in the neutron--neutron potential \cite{Berryman:2022zic}.

Axions and light BSM scalars may be copiously produced by nucleon bremsstrahlung processes in the merger environment and subsequently free-stream through the dense nuclear matter or be trapped \cite{Harris:2020qim,Dev:2021kje}.  In the free-streaming case, these particles provide an extra channel to cool down the merger remnant; while this cooling only has minor changes on the dynamics of the merger at the level of the GW simulations conducted in \cite{Dietrich:2019shr}, it should be noted that the emitted particles may subsequently decay to produce exotic photon signals \cite{Diamond:2021ekg}. In the  trapped regime, scalars  can contribute a larger thermal conductivity than the trapped neutrinos in some parts of the parameter space \cite{Dev:2021kje}, thereby leading to faster thermal equilibration than expected. Future observations of the early post-merger phase of a neutron star merger could effectively probe a unique range of the scalar parameter space. Numerical relativity simulations of the post-merger phase predict a transient GW  signal on dynamical timescales $\sim$ tens of milliseconds,  with complex  morphology and a characteristic peak frequency in the range  2-4 kHz. Observations of this phase  may be possible with third generation detectors like Einstein Telescope \cite{Punturo_2010} and Cosmic Explorer~\cite{Evans:2021gyd} or new proposals such as NEMO targeting the kHz range \cite{Ackley:2020atn}.

Finally, now that GW detectors have assembled a sizable catalog of events, it is possible to probe new physics and dark sectors via the parameter distributions of the detected objects. This includes detailed statistical analyses of the black hole mass and spin distribution testing possible new physics origins of these compact objects \cite[see e.g.][]{Bramante:2017ulk,Fernandez:2019kyb,DeLuca:2021wjr}. Additionally, since standard astrophysical arguments give rise to restrictions on the mass spectrum of compact objects, the detection of even a small number of objects in particular mass ranges can point to beyond--Standard-Model physics. GW observations to date have already yielded at least one such surprise: one of the component masses of GW190521 lies within the `pair-instability mass gap'~\cite{LIGOScientific:2020iuh}, a regime in which the formation of black holes by standard astrophysical mechanisms is challenging. Such objects are readily accommodated by new physics that modifies stellar evolution. In particular, new weakly-coupled light particles (such as axions or dark photons) would affect the cooling rates of stars, perhaps allowing for the astrophysical formation of BHs in mass ranges which would not otherwise be possible \cite[see e.g. subsection \ref{sec:BHPS} and][]{Croon:2020ehi,Croon:2020oga,Sakstein:2020axg,Sakstein:2022tby}.

Further, in dissipative dark matter scenarios, black holes with exotic masses may form directly from dark matter that cools through dark radiative processes at $z\lesssim 30$. In that case, the dark sector contains both light and heavy particles, and the black hole mass is proportional to the coldest temperature the gas can reach \cite{low/lynden-bell:1976,rees:1976}. This is possible if, for example, dark matter consists of two fundamental fermions oppositely charged under a dark $U(1)$, mediated by a massless photon. Such a dark sector has all of the usual cooling processes of hydrogen gas \cite{Rosenberg:2017qia,Ryan:2021dis}. Early estimates of the black hole population produced in this scenario \cite{Shandera:2018xkn} found a significant population of sub-solar mass dark black holes may be produced when the heavier dark fermion has a mass above that of the standard model proton. The black hole population observed in mergers has been used to bound this scenario under the assumption that none of the black holes have a dark matter origin or that the unusual low-mass event GW190425 \cite{LIGOScientific:2020aai} was a dark black hole merger \cite{Singh:2020wiq}. The null subsolar mass search results also constrain dissipative models \cite{LIGOScientific:2018glc, LIGOScientific:2019kan,LIGOScientific:2021job}. Recent work on the molecular chemistry of the dark gas \cite{Ryan:2021dis,Gurian:2021qhk,Ryan:2021tgw} will allow the estimated population of \cite{Shandera:2018xkn} to be revised using numerical studies of the cooling of a dark hydrogen gas. Other scenarios for black hole formation from dark matter, including super-massive black holes, were considered in \cite{DAmico2017,Latif:2018kqv,Choquette2019,Chang2019}. Dark matter may form other compact objects as well. Dark neutron stars were considered in \cite{Hippert:2021fch}, and dark white dwarfs in \cite{Ryan:2022hku}. All of these objects present potential targets for future GW observations.

\subsection{Formation of black holes from dark matter capture in neutron stars}
\label{subsec:implosion}


Neutron stars in binaries could capture certain types of dark matter particles that thermalise and collapse to form mini black holes \cite{Bertoni:2013bsa}. For example, in the asymmetric dark matter scenario \cite{McDermott:2011jp, Zurek:2013wia} dark matter particles do not self-annihilate due to the assumed asymmetry between the number density of particles and antiparticles. In the case of bosonic dark matter, the Chandrasekhar limit is much greater than that for fermions \cite{Shapiro:1983du}. Consequently,  bosonic dark matter could undergo gravitational collapse sooner than fermionic dark matter.  Eventually, accretion of such dark matter particles could lead to the implosion of one or both the companion neutron stars to black holes before they merge \cite{McDermott:2011jp, Bramante:2014zca}. Thus, the universe might contain three distinct populations of compact binary mergers in the mass range $\sim \mbox{1--3}\, M_\odot$: one containing only neutron stars, a second population of only black holes, and a third consisting of a neutron star and a black hole. The mixed binary population of neutron stars and black holes, however, is not expected to be significant since either both neutron stars will implode before they merge or neither would. 

The relative fractions of the different populations depend on the implosion time-scale $t_c$ and the delay-time $t_d$ between the formation of neutron stars in a binary and their merger \cite{Singh:2022abc}. If typical implosion time scales are larger than typical delay times then implosions will be rare and the binary black hole merger population in this mass range will be small. On the other hand, if the collapse time scales are small compared to delay times then the Universe might have a significant fraction of binary black hole mergers in this mass range. Currently, neither of these time scales is known very well but gravitational-wave observations might determine both. 

Future gravitational-wave detector networks, including upgrades \cite{KAGRA:2013rdx} of Advanced LIGO \cite{Harry:2010zz}, and Virgo \cite{VIRGO:2014yos} and new facilities such as the Cosmic Explorer \cite{Evans:2021gyd} and Einstein Telescope \cite{Punturo_2010}, can discriminate between the different populations of compact binaries in the mass range of neutron stars by measuring their effective tidal deformability $\Tilde{\Lambda}$ \cite{Hinderer:2007mb}. The effective tidal deformability of binary black holes is zero \cite{LeTiec:2020bos}, while that of binary neutron stars is nonzero \cite{Hinderer:2007mb}. Thus, gravitational-wave observations can infer the relative fractions of the different merger populations and hence the distribution of the collapse time for the neutron stars to implode to black holes. The implosion time-scale can in turn be used to constrain a combination of the mass, interaction cross-section and the local density of dark matter particles. Thus, from the relative fraction of the observed population of compact binaries in the $\mbox{1–3} M_\odot,$ gravitational wave detectors can set limits or constrain the properties of dark matter \cite{Singh:2022abc}.

\subsection{Formation of (sub)solar mass black holes from DM capture inside compact objects}
\label{sec:dmicocollapse}


Accumulation of DM inside compact astrophysical objects via interactions between DM and constituents of compact astrophysical objects have been studied for some time\,\cite{Goldman:1989nd,Gould:1989gw}.  For large regions of well-motivated particle DM parameter space, it is possible that the accumulation of DM particles will result in a sufficient bulk of DM such that the accumulated DM particles collapse to form a black hole at the center of a star or planet. Essentially all observatories with sensitivity to the dynamics and properties of compact objects like neutron stars and black holes can be utilized for this DM detection strategy. This includes upcoming pulsar searches at FAST~\cite{FAST2011}, SKA \cite{SKA2008}, and CHIME~\cite{CHIME2021}, gravitational wave observations at detectors like LIGO/Virgo/KAGRA~\cite{LIGOScientific:2019kan, LIGOScientific:2021job, Nitz:2021vqh}, along with associated optical transient searches for NS(/BH) mergers at wide-field observatories like BlackGEM, LS4, and the Zwicky Transient Factory \cite{Ghosh:2014yga,Nugent:2020,ZTF2014}. 

A small BH formed from DM at the center of a star grows by accretion of stellar material, and can consume the entire star in this process\,\cite{Kouvaris:2007ay, Bertone:2007ae, deLavallaz:2010wp, Kouvaris:2010vv, McDermott:2011jp, Kouvaris:2011fi, Kouvaris:2011gb, Bramante:2013hn, Bell:2013xk, Bramante:2014zca, Bramante:2015cua, Bramante:2016mzo, Bramante:2017ulk, NSvIR:GaraniGenoliniHambye, Kouvaris:2018wnh, Kopp:2018jom, Acevedo:2019gre, Janish:2019nkk, East:2019dxt,Tsai:2020hpi,Takhistov:2020vxs,Dasgupta:2020mqg,Acevedo:2020gro, NSvIR:Bell2020improved,NSvIR:Bell:ImprovedLepton, Acevedo:2020avd,NSvIR:anzuiniBell2021improved, Garani:2021gvc,Giffin:2021kgb}. The DM models discover-able with this scenario include both bosonic and fermionic asymmetric DM with masses in the range $\sim {\rm keV - 10^{10} ~GeV}$, and for composite DM this mass range extends further to $\sim 10^{45}$ GeV. The processes relevant for BH formation in stellar interiors include capture of DM in the star, a period of DM thermalization within the stellar interior, collapse of the collected DM inside the star, and growth of the resulting small BH, which depends upon the rate for BH accretion of nearby material and Hawking radiation. While compact stars have excellent sensitivity to a variety of asymmetric DM models, it has also been shown that main sequence stars and planets like the Sun and Earth \cite{Acevedo:2020gro}, along with high redshift Pop III stars \cite{Ellis:2021ztw}, can also be used to search for DM through the formation of BHs in their interiors, including IceCube searches for neutrinos \cite{Acevedo:2020gro}, along with JWST and EELT searches for Pop III stars \cite{JWST:Gardner:2006ky,ELT:neichel2018overview}.

If DM destroys a compact star in the process detailed above, the resultant BH has a mass which is very similar to that of the compact astrophysical object\,\cite{Bramante:2017ulk, Kouvaris:2018wnh, East:2019dxt}.  This process is most efficient in dense stellar objects like neutron stars and white dwarfs, since these compact object result in collected DM forming a smaller, gravitationally bound clump inside the dense stellar interior of a compact star. In the case of DM that is captured and forms a small core within a WD, the resulting collapse of the DM core can cause a white dwarf supernova explosion either prior to or after the formation of a small BH made of DM \cite{Bramante:2015cua,Acevedo:2019gre,Janish:2019nkk,Acevedo:2020avd}. In the case of NSs, there are many signatures of NSs which have been converted to BHs, including an absence of pulsars near the centers of galaxies, an associated ``pulsar maximum age curve" extending outward from the galactic center \cite{Bramante:2014zca,Bramante:2015dfa}, the distribution of Fast Radio Bursts, NS merger kilonovae, and resulting r-process metals in galaxies \cite{Bramante:2016mzo}, and an unexpected population of around solar mass black holes in galaxies \cite{Bramante:2017ulk}. Since the masses of objects formed from DM-induced collapse of stars are between $\sim$ 0.2 $M_\odot$ and 3 $M_\odot$\,\cite{Ozel:2016oaf, 2018MNRAS.480.4505J}, this implies that the masses of these BHs, sometimes called ``transmuted black holes" (TBHs, \cite{Takhistov:2017bpt}) are in the above mentioned range. 

In other motivated theories, DM can be composed of asteroid-mass PBHs, formed through variety of mechanism in the early Universe (e.g.,~\cite{Cotner:2018vug,Cotner:2019ykd,Kusenko:2020pcg,Deng:2017uwc}). Such PBHs can be efficiently captured by compact objects, white dwarfs and neutron stars~\cite{Capela:2013yf}. Captured PBHs will settle within and devour the host stars. Ejection of neutron-rich material from neutron stars being consumed by PBHs sets a favorable environment for $r$-process nucleosynthesis, and such events can help address the origin of heavy elements such as gold~\cite{Fuller:2017uyd}. More so, the destruction of neutron stars by small PBHs, contributing to the DM, is consistent with the under-abundance of observed pulsars in the Galactic Center~\cite{Fuller:2017uyd}.
A multitude of astrophysical signatures accompany these violent events, including 
fast radio bursts~\cite{Fuller:2017uyd,Abramowicz:2017zbp,Kainulainen:2021rbg}, 511 keV radiation~\cite{Fuller:2017uyd}, formation of a new class of microqusars~\cite{Takhistov:2017nmt} as well as ``orphan'' gamma-ray bursts~\cite{Takhistov:2017nmt} and kilonovae~\cite{Fuller:2017uyd,Bramante:2017ulk} without accompanying strong gravitational wave emission.
Devoured compact objects, neutron stars~\cite{Takhistov:2017bpt,Bramante:2017ulk} as well as white dwarfs~\cite{Takhistov:2017bpt}, will leave behind $\sim 0.2-3 M_{\odot}$ TBH remnants. Such (sub)solar-mass TBHs are distinct from (sub)solar-mass PBHs, which were formed in the early Universe prior to star formation.

BHs with mass $\lesssim 2.5 M_{\odot}$ are not expected from conventional stellar evolution due to neutron star Tolman–Oppenheimer–Volkoff stability limit (e.g.~\cite{Godzieba:2020tjn}), analogous to the Chandrasekhar limit for white dwarfs. Thus, detection of such (sub)solar-mass BHs constitutes a smoking gun and attractive target for new physics. Gravitational wave detectors are sensitive to sub-solar mass BHs\,\cite{LIGOScientific:2019kan, LIGOScientific:2021job, Nitz:2021vqh}.
However, discrimination with GWs between compact objects of the same mass, such as a BH and a neutron star, is challenging without electromagnetic counterparts and determination of higher order GW effects. TBHs can contribute to a variety of merging binary events~\cite{Takhistov:2017bpt,Bramante:2017ulk}. In fact, one can consider such possibility for the first and only confirmed multi-messenger merger event, GW170817 \cite{LIGOScientific:2017vwq}. The excellent environmental information of the host galaxy of GW170817, NGC 4993, provides an opportunity for detailed considerations of such scenario \cite{Tsai:2020hpi}, including the DM and NS distribution. These scenarios are distinct from PBH binary formations, PBH-PBH~\cite{Sasaki:2016jop,Bird:2016dcv,Clesse:2016vqa} and PBH-NS~\cite{Sasaki:2021iuc} systems.

Recent LIGO/Virgo/KAGRA GW observations of binary merger events provide some guidance for how to identify the origin of detected (sub)solar-mass BHs. In particular, the observation of GW190425~\cite{LIGOScientific:2020aai} and GW190814~\cite{LIGOScientific:2020zkf}, consistent with BH masses in the range $\sim 1.5-2.6 M_{\odot}$, highlight the capabilities of these observatories. TBHs formed as remnants of devoured compact objects by DM will have the original compact object mass distribution, which is known from astrophysics such as supernovae models. A reliable statistical method for distinguishing the origin of such BHs based on BH mass-function has been put forth in~\cite{Takhistov:2020vxs}, which demonstrated that BH events with mass $\gtrsim 1.5 M_{\odot}$ are unlikely to stem from neutron star implosions. Employing this method, data analyses
of upcoming GW observations will be able to distinguish between solar-mass BHs and NSs with high
confidence.  Merger rates as a function of redshift constitute another important handle to distinguish between PBHs and TBHs in identifying the origin of detected BH events. Using an analytical calculation of the merger rate in a simplified scenario, ref.\,\cite{Dasgupta:2020mqg} showed that the merger rates at various redshifts for the TBH-TBH binaries depend on the underlying particle physics parameters (which are responsible for the formation of the TBH) and is completely different compared to merger rate v/s redshift distribution of PBHs.  This distinct redshift distribution can be understood easily: TBHs are formed from neutron stars or white dwarfs which form at much later times compared to PBHs.  It was shown that, depending on the particle physics parameter space, near future gravitational wave telescopes, like Einstein Telescope and Cosmic Explorer, can detect TBH mergers at large rates.  One by-product of this discovery will be that we can probe new parts of the DM - Standard Model interaction strength parameter space, which cannot be probed by laboratory experiments\,\cite{Dasgupta:2020mqg}.

The above calls for further detailed studies across a multitude of complimentary directions, including stellar evolution, mechanisms of formation of astrophysical black holes, emission, and nucleosynthesis associated with compact objects and their dynamics as well as reinvigorated observational campaigns for distinct and multi-messenger signatures.

\medskip

\section{Conclusion}
Over the last decade, astroparticle physics has become a rich research field through which a variety of dark matter candidates can be probed. In this white paper, we have focused on opportunities relating to extreme astrophysical environments, including the cores of stars, the neighborhoods of compact objects such as black holes and neutron stars, compact object binary mergers, and supernova explosions.

One class of astrophysical opportunities for particle physics relates to new particles which couple to electrons, nucleons, and photons which are present in extreme environments.
The high temperatures and densities in these environments are far beyond what can be created in lab-based experiments, allowing enhanced production of and increased sensitivity to weakly-coupled new particles. 

An entirely different class of opportunities  relies only on the gravitational interaction of the new particles, and can be used to search for new sectors with arbitrarily feeble SM couplings. This includes black hole superradiance searches for ultralight bosons, which leads to BH spindown and GW emission. Other directions include imprints on gravitational wave-forms left by concentrated environments of dark matter or by mergers of exotic compact such as boson stars. 

The next decade will see continued development in this field. Theoretical understanding of the underlying astrophysical processes and the new particle production mechanisms are rapidly improving.  Experimental progress is expected in  observations across the EM spectrum (including radio, optical, X-ray, and gamma ray observation), observations of neutrinos from astrophysical objects, and ever-increasing sensitivity and broadening frequency coverage of gravitational wave experiments. The development of multi-messenger astronomy gives rise to the tantalising prospect that multiple scenarios can be probed concurrently.

\section*{Acknowledgments}

BJK thanks the Spanish Agencia Estatal de Investigaci\'on (AEI, MICIU) for support to the Unidad de Excelencia Mar\'ia de Maeztu Instituto de F\'isica de Cantabria, ref. MDM-2017-0765.
E.B. is supported by NSF Grants No. PHY-1912550 and AST-2006538, NASA ATP Grants No. 17-ATP17-0225 and 19-ATP19-0051, NSF-XSEDE Grant No. PHY-090003, and NSF Grant PHY-20043. 
R.B. acknowledges financial support provided by FCT -- Funda\c{c}\~{a}o para a Ci\^{e}ncia e a Tecnologia, I.P., under the Scientific Employment Stimulus -- Individual Call -- 2020.00470.CEECIND.
N.R. is supported by the Natural Sciences and Engineering Research Council of Canada.
EV was supported in
part by the US Department of Energy (DOE) Grant DE-
SC0009937. 
Research at Perimeter Institute is supported in part by the Government of Canada through the Department of Innovation, Science and Economic Development Canada and by the Province of Ontario through the Ministry of Economic Development, Job Creation and Trade. This research was undertaken thanks in part to funding from the Canada First Research Excellence Fund through the Arthur B. McDonald Canadian Astroparticle Physics Research Institute.


\bibliographystyle{JHEP.bst}
\bibliography{main.bib}

\end{document}